\documentclass[%
 reprint,
 amsmath,amssymb,
 aps,
]{revtex4-2}
\usepackage{graphicx} 
\usepackage[dvipsnames]{xcolor}
\usepackage{ulem}
\usepackage{dcolumn}
\usepackage{bm}
\usepackage{graphicx}       

\newcommand{\var}{\mathrm{var}}
\newcommand{\covar}{\mathrm{covar}}

\begin{document}

\preprint{APS/123-QED}

\title{Teleportation-based squeezer for bosonic cluster states}


\author{Michal Matulík}
\email{matulik@optics.upol.cz}

\author{Radim Filip}%
\email{filip@optics.upol.cz}
\author{Petr Marek}
\email{marek@optics.upol.cz}

\affiliation{%
 Department of Optics, Palacký University, 17. listopadu 1192/12,  771 46 Olomouc, Czech Republic
}%
\begin{abstract}
The one-way quantum computation utilizing bosonic modes of light offers unmatched scalability of light modes, and it has seen rapid experimental development recently. Scalability requires robust and low-error gates and measurements. Squeezing gate is one of the necessary Gaussian operations. We find the optimal squeezing gate in cluster state architecture. Our approach newly uses amplitude transmission coefficients of unbalanced beam splitters and homodyne detection with subsequent unity-gain feed-forward to squeeze the input state. The approach outperforms the current method based on optimally rotated homodyne detection but fixed balanced beam splitters. The performance of both cluster state squeezers is evaluated for Gaussian and non-Gaussian input states. We use different metrics to benchmark the quality of squeezed output states. The result opens a road to low-noise squeezing gates in experimentally achievable cluster states.

\end{abstract}

\maketitle
\section{Introduction}
Quantum computing takes advantage of the inherent properties of quantum systems to find solutions to problems untractable by classical computation \cite{ladd,harrow}. Many platforms are investigated for achieving the goal \cite{arute,ringbauer,loic,qiang}. Among them, the one-way computation \cite{raussendorf,briegel} utilizing bosonic modes of light offers unparalleled scalability of the number of modes \cite{zhang,gu,yokohama,larsen,asavanant}. The actual processing of bosonic modes requires two fundamental kinds of quantum resources.

The first resource is the non-Gaussian states, used for encoding the information \cite{gkp,cat}, error correction \cite{fukui}, and as a resource for non-Clifford gates \cite{gkp,lloyd}. Non-Gaussian states are defined as those states that cannot be expressed as mixtures of Gaussian states. Negativity in the Wigner function is sufficient to claim non-Gaussian nature, and it is also a necessary condition for achieving universality in quantum computing \cite{mari}. Preparation of non-Gaussian states for traveling modes of light is a challenging task that is steadily being addressed by tailored optical circuits employing photon number resolving detectors for probabilistic generation \cite{tiedau,motes}. Hand in hand with the advancement of state preparation techniques follows the design of novel ways to detect and verify the non-Gaussian nature of the states \cite{lachman,fiurasek}.

The second resource, required for efficient quantum computation with bosonic modes of light, encompasses the set of Gaussian operations necessary for the realization of the Clifford gates \cite{bartlett}. In the resource theory of non-Gaussianity \cite{walschaers}, Gaussian operations are considered to be `free', but in practical scenarios they have associated costs in the form of loss and added noise \cite{yoshikawa,kashiwazaki}. The quintessential example of such added noise can be found in quantum teleportation \cite{bennett,braunstein_kimble,braunstein_loock,furusawa} - the fundamental Gaussian channel representing the basic way to propagate a quantum state through the optical cluster. The complete set of Gaussian operations can be simplified to phase shift, displacement, beam splitter, and squeezing, which together can be used to create an arbitrary Gaussian operation \cite{braunstein_squeezing}. The first three of the four listed operations require only linear optics and can be readily incorporated in the cluster without incorporating significant losses \cite{larsen,asavanant}.  The squeezing, however, is an active operation that needs to be realized either in an active medium \cite{andersen} and then injected into a cluster state, or by a measurement-induced procedure \cite{filip_marek} directly in a cluster state. In both cases, it is impossible to achieve infinite squeezing, which results in undesired decoherence.

In this paper, we study several ways in which squeezing can be implemented as a part of the teleportation protocol in cluster-state architecture. We compare the different approaches and optimize over the free parameters of the circuits for several levels of finite resources and realistic imperfections. We evaluate the performance of the measurement-induced squeezing for Gaussian and non-Gaussian states, for which we study how their non-Gaussian nature diminishes during the protocol.
We are focused on the non-Gaussian nature, rather than non-classicality of the states, because the squeezing operation is, in itself, non-classical. Non-classical states are those that cannot be expressed as mixtures of coherent states~\cite{mandel}, and squeezing does produce them. Measures of non-classicality based on the P-function do not yield relevant information; we use the Wigner function description instead. 

In Section \ref{sec:tel_sq}, we describe several ways to perform teleportation-based squeezing and use Heisenberg formalism to derive the formulas for noise added during the protocols and the expressions for characteristic and Wigner functions of the transformed states. We also discuss the metrics used for the evaluation of the test states - the vacuum and the single photon state. In Section \ref{sec:analysis}, we then show the numerical analysis results for the various protocols and test states, focusing on preserving non-Gaussian features of the transformed quantum states. Section \ref{sec:conclusion} then concludes the manuscript.

\begin{figure}[hpbt]
\includegraphics[width=0.5
\textwidth]{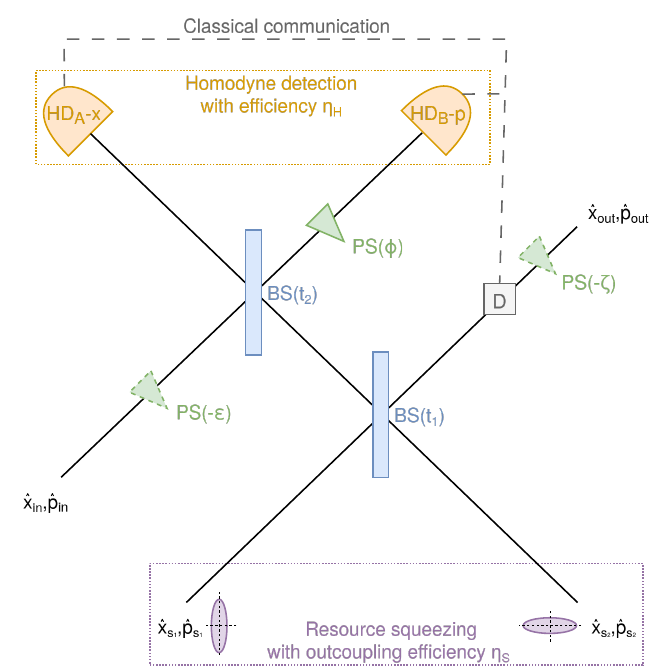}
\caption{\label{fig:scheme} Quantum                                                                                                                                                                                                                                                                                                                                                                                                                                              circuit of teleportation-based squeezer. \text{HD$_{\text{A}}$} - homodyne measurement in $x$-quadrature, \text{HD$_{\text{B}}$} - homodyne measurement in $p$-quadrature , \text{BS}($t_1$) - beam splitter with transmission coefficient $t_1$, for generation of TMSV state. \text{BS}($t_2$) with transmission coefficient $t_2$ mixes the input state with one mode of the TMSV. \textit{D} - feed-forward displacement, displaces the second mode of TMSV according to homodyne measurements. The gain of feed-forward is always set to unity. The gray dashed line indicates classical communication.  The quadratures $\hat{x}_{in}$, $\hat{p}_{in}$ and $\hat{x}_{out}$, $\hat{p}_{out}$ describe the input and output states, respectively. The quadratures $\hat{x}_{s_1}$, $\hat{p}_{s_1}$ and $\hat{x}_{s_2}$, $\hat{p}_{s_2}$ correspond to oppositely squeezed vacuum states. The overall homodyne detection efficiency is denoted as $\eta_H$ and includes also mode-matching at the BS($t_2$). The excess noise in the anti-squeezed quadrature in the preparation of the squeezing resource is considered due to the overall outcoupling efficiency of squeezers, denoted as $\eta_S$. It effectively also includes the mode-matching at the BS($t_1$). The phase shift \text{PS}($\phi$) adjusts the phase of measured mode B and enables squeezing in output mode if $\phi \neq 0$. Beam splitters \text{BS}($t_i$) with variable transmission coefficient $t_i$ also contribute to squeezing in the output mode if $t_i^2 \neq 1/2$.  The auxiliary phase shift PS($-\zeta$) and PS$(-\epsilon)$ compensate the rotation caused by PS($\phi$).}
\end{figure} 

\section{Teleportation-based squeezer}\label{sec:tel_sq}
A field of traveling light can be modeled by a harmonic oscillator described by quadrature operators $\hat{x}$ and $\hat{p}$ with $[\hat{x},\hat{p} ] =  i$. We aim to implement phase-sensitive amplification described by the following input-output relations 
\begin{equation}\label{eq:squeezing_id}
    \hat{x}_{out} = s \hat{x}_{in},\quad \hat{p}_{out} = \frac{1}{s}\hat{p}_{in},
\end{equation}
where $s>0$ is the squeezing parameter. The parametric amplification can synthesize squeezed states \cite{yuen} with uncertainties below the shot noise limit in one quadrature, at the expense of higher uncertainty in the other quadrature. Equation \eqref{eq:squeezing_id} describes the squeezing operation. A squeezer is a device that creates squeezed states.  To implement the squeezing operation in a measurement-based quantum computation using cluster states, we can incorporate the squeezing into a single step of a teleportation protocol \cite{menicucci_1}. Such a generalized quantum teleportation protocol is schematically depicted in Fig.~\ref{fig:scheme}. The quantum resource of the teleportation protocol is a two-mode squeezed vacuum state (TMSV) that is created by mixing two orthogonally squeezed vacuum states on a beam splitter with transmission coefficient $t_1$ \cite{EPR}. The first mode of this entangled state then interacts with the input state, described by quadratures $\hat{x}_{in}$ and $\hat{p}_{in}$, on another beam splitter with transmission coefficient $t_2$. The resulting modes $A$ and $B$ are measured by a pair of homodyne detectors generating classical signals that drive displacement feed-forward operation on the second mode of the resource TMSV state. The gain in the feed-forward is always set to unity. The homodyne detector HD$_{\text{A}}$ measures fixed quadrature $\hat{x}_A$. The homodyne detector HD$_{\text{B}}$ measures $\hat{p}_B$. The generalized scheme also involves three phase shift gates with parameters $\phi$, $\zeta$, and $\epsilon$.  In a regular teleportation protocol, it holds $t_1 = t_2 = \frac{1}{\sqrt{2}}$ and $\phi = \zeta= \epsilon = 0$. When $\phi$ is non-zero then HD$_{\text{B}}$ measures rotated quadrature $\hat{p}_B \cos\phi + \hat{x}_B \sin{\phi}$.

Quantum circuit in Fig.~\ref{fig:scheme} can be modified to realize the squeezing operation in two ways. The approach typical for the cluster states keeps $t_1 = t_2 = \frac{1}{\sqrt{2}}$ but allows for non-zero values of $\phi$, $\zeta$, and $\epsilon$, see Appendix \ref{sec:appB}. We are going to call it the phase-shift squeezer (PS-sq). Having a non-zero value of $\phi$ effectively implements the shear gate \cite{menicucci_1,menicucci_2,menicucci_3,menicucci_4,menicucci_5} which can be converted to the desired axis-aligned squeezing by properly adjusting the correcting phases $\zeta$ and $\epsilon$ \cite{weedbrook,kalajdzievski}, together with the feed-forward operation, see Appendix \ref{sec:appB}. For the case of an ideal scenario with an infinitely squeezed resource state, the realized transformation is described by Eq. \eqref{eq:squeezing_id} with the squeezing parameter
\begin{equation}\label{eq:s_ps}
    s_{PS} = \sqrt{\frac{1}{2} \big (2+ l^2-l \sqrt{4 + l^2} \big )},
\end{equation}
where
\begin{equation}\label{eq:k}
l = 2 \tan{\phi}.
\end{equation}
We can see that the PS-squeezer can realize arbitrary squeezing with $s_{PS} \in (0,1]$. The orthogonal squeezing could be easily realized by applying a $\pi/2$ phase shift to the state.

The second approach towards teleportation-based squeezer keeps the phases equal to zero, but varies the transmission coefficients $t_1$ and $t_2$ of the used beam splitters. We will call this approach beam splitter-based squeezer (BS-sq). Again, in the lossless case, the operation realizes the squeezing gate \eqref{eq:squeezing_id} with the squeezing parameter
\begin{equation}\label{eq:s_bs}
    s_{BS} = \frac{r_1 r_2}{t_1 t_2}.
\end{equation}
Note that, unlike the PS-sq scheme, the same value of $s_{BS}$ can be reached by different values of $t_1$ and $t_2$, which opens up the avenue for optimization. We maximize the fidelity between the output state and the squeezed state for various resources, efficiencies, and input states. The minimization with respect to total added noise is discussed in Appendix \ref{sec:appE}. We also discuss the optimization with respect to the minimum of the Wigner function later in the text. The coefficients $t_1$ and $t_2$ are, from now on, called the optimization parameters.

Finally, both approaches can be combined into a single general scheme, BSPS-sq, that applies both squeezing and shear operations. After suitable phase corrections $\zeta$ and $\epsilon$, which are generally different than those in the case of PS-sq, the squeezing-gate parameter $s$ in Eq. \eqref{eq:squeezing_id} can be found to be, see Appendix \ref{sec:appC}:
\begin{equation} \label{eq:s}
\begin{split}
    s_{BSPS} &= \frac{1}{g\sqrt{2}}\bigg(1+ g^4+k^2 \\ &- \sqrt{(1-g^4)^2 + 2 (1+g^4)k^2 + k^4 }\bigg)^{1/2},
\end{split}
\end{equation}
where
\begin{equation}\label{eq:g and k}
    g = \frac{r_1 r_2}{t_1 t_2}, \; \; \;  \; k = \frac{1}{t_2^2} \tan{\phi}.
\end{equation}
Again, please note that the same value of $s_{BSPS}$ can be reached by different values of optimization parameters $t_1$, $t_2$, and $\phi$. The PS-sq is a special case of the BSPS-sq when $g=1$. The BS-sq can also be obtained from BSPS-sq when $k = 0$, then $s_{BSPS} = g $.

\begin{table}
\caption{\label{tab:table1}Overview of types of squeezers, corresponding squeezing parameters, and optimization parameters. In the case of PS, there is no need for optimization because of only one free parameter $\phi$.}
\begin{ruledtabular}
\begin{tabular}{lccc}
\textrm{Teleportation squeezer:}&
\textrm{PS}&
\textrm{BS}&
\textrm{BSPS}\\
\textrm{Squeezing parameter $s$:}&$s_{PS}$& $s_{BS}$& $s_{PSBS}$\\
\textrm{Optimization parameters:}&$\phi$& $t_1,t_2$& $\phi,t_1,t_2$\\
\end{tabular}
\end{ruledtabular}
\end{table}

The infinitely squeezed vacuum states in modes $s_1$ and $s_2$ used to create the entangled resource are an idealization useful for showing the action of the gate, but they are nonphysical. In realistic scenarios, they need to be replaced by states with finite squeezing and purity. As is illustrated in Fig.~\ref{fig:scheme}, such states can be modeled as pure squeezed states with variances $\var (\hat{x}_{sj}) $ and $\var (\hat{p}_{sj}) = \big (4 \var (\hat{x}_{sj}) \big)^{-1} $ that are also subject to lossy channels with intensity transmissivity $\eta_S$. The channels can be modeled by beam splitters with the transmission coefficients $\sqrt{\eta_S}$ and vacuum in the idle mode. In the context of teleportation, the channel adds excess noise into the quadratures of the teleported state. The variances are often expressed in decibels relative to the variance of the input state. In current-day experiments, 10-15 dB of squeezing before the lossy channel can be considered available \cite{schanbel,xanadu}.  Realistic homodyne detectors with limited quantum efficiency $\eta_H$ are another source of imperfections. Such imperfections, too, can be simulated as variable beam splitters with transmission coefficients $\sqrt{\eta_H}$ placed just before homodyne detectors HD$_\text{A}$ and HD$_\text{B}$.

When we take into account all the imperfections discussed above, the quadrature operators of the transformed state have the following form:
\begin{equation}\label{eq:xtel,ptel}
\begin{split}
\hat{x}_{out} &= s \hat{x}_{in} + \hat{N}_{x_{out}}, \\
\hat{p}_{out} &= \frac{1}{s} \hat{p}_{in} + \hat{N}_{p_{out}},
\end{split}
\end{equation}
The general squeezing parameter $s$ stands for the individual squeezing parameters $s_{BS}$, $s_{PS}$, and $s_{BSPS}$ described above and as summarized in Table \ref{tab:table1}.
The operation still transforms mean values ideally \eqref{eq:squeezing_id}, but now it also contains noise operators $\hat{N}_{x_{out}}$ and $\hat{N}_{p_{out}}$. They can be calculated from the rotated noise operators $\hat{N}^{(\zeta)}_{x_{out}}$ and $\hat{N}^{(\zeta)}_{p_{out}}$:
\begin{equation}
    \begin{split}
        \hat{N}_{x_{out}} &= \hat{N}^{(\zeta)}_{x_{out}} \sqrt{\frac{1-g^2s^2}{1 -s^4}} +  \hat{N}^{(\zeta)}_{p_{out}} \sqrt{\frac{g^2s^2-s^4}{1 -s^4}}, \\
        \hat{N}_{p_{out}} &= \hat{N}^{(\zeta)}_{p_{out}} \sqrt{\frac{1-g^2s^2}{1 -s^4}} - \hat{N}^{(\zeta)}_{x_{out}}\sqrt{\frac{g^2s^2-s^4}{1 -s^4}}. \\
    \end{split}
\end{equation}
The fractions in the upper equation are obtained from the trigonometric functions $\cos\zeta$ and $\sin\zeta$, which are a consequence of the action of auxiliary phase rotation PS($-\zeta$).
The rotated noise operators represent the expression of quadrature operators:
\begin{equation}\label{eq:N'}
\begin{split}
    \hat{N}^{(\zeta)}_{x_{out}} &= \frac{1}{t_1}  \big( \sqrt{\eta_S}\hat{x}_{s_1} + \sqrt{1-\eta_S} \hat{x}_{\tilde{s_1}}^0\big)\\   & +\sqrt{\frac{1-\eta_H}{\eta_H}} \frac{g}{r_2} \hat{x}^0_{\tilde{A}}, \\
    \hat{N}^{(\zeta)}_{p_{out}} &=  - \frac{1}{r_1} \big( \sqrt{\eta_S}\hat{p}_{s_2} + \sqrt{1 - \eta_S} \hat{p}_{\tilde{s_2}}^0\big)\\ &+ \sqrt{\frac{1-\eta_H}{\eta_H}} \frac{1}{g}\bigg \{  \frac{1}{ t_2} \hat{p}^0_{\tilde{B}}  +  k \bigg({ t_2} \hat{x}^0_{\tilde{B}}   + {r_2}\hat{x}^0_{\tilde{A}} \bigg) \bigg \}.
\end{split}
\end{equation}
The modes denoted with a tilde are the auxiliary vacuum modes simulating the added noise from the imperfect detectors and imperfect preparation of the squeezed resource state. 
To quantify the added noise, it is appropriate to find the variances of the output quadratures $\text{var}(\hat{x}_{out})$ and $\text{var}(\hat{p}_{out})$.
The noise operators commute with the input quadratures. Thus, for the variance, it holds $\text{var}(\hat{x}_{out}) = s^2 \text{var}(\hat{x}_{in}) + \text{var}(\hat{N}_{x_{out}}) $, similarly for $\text{var}(\hat{p}_{out})$.
The noise variances $\var (\hat{N}_{x_{out}})$ and $\var(\hat{N}_{p_{out}})$ are equal to
\begin{equation}\label{eq:varNx}
\begin{split}
\var (\hat{N}_{x_{out}}) &= \frac{(1-g^2 s^2)}{1-s^4} \var{(\hat{N}^{(\zeta)}_{x_{out}})}\\  &+ \frac{(g^2 s^2- s^4)}{1-s^4} \var{(\hat{N}^{(\zeta)}_{p_{out}})}, \\
\var (\hat {N}_{p_{out}}) &= \frac{(1-g^2 s^2)}{1-s^4} \var{(\hat{N}^{(\zeta)}_{p_{out}})} \\ & +\frac{(g^2 s^2- s^4)}{1-s^4} \var{(\hat{N}^{(\zeta)}_{x_{out}})},
\end{split}
\end{equation}
where the rotated noise variances are $\var{(\hat{N}^{(\zeta)}_{x_{out}})}$ and $\var{(\hat{N}^{(\zeta)}_{p_{out}})}$ are
\begin{equation}\label{eq:N1N2}
    \begin{split}
        \var{(\hat{N}^{(\zeta)}_{x_{out}})}  & = \frac{1}{t_1^2} \big ( \var(\hat{x}_{s_1}) \eta_S + \frac{1}{2}(1-\eta_S ) \big )\\ 
        &+  \frac{1-\eta_H}{\eta_H} \frac{g^2}{r_2^2} \frac{1}{2}, \\
        \var(\hat{N}^{(\zeta)}_{p_{out}}) & = \frac{1}{r_1^2} \big ( \var(   \hat{p}_{s_2}) \eta_S + \frac{1}{2}(1-\eta_S ) \big )\\ &+  \frac{1-\eta_H}{\eta_H} \frac{1}{g^2} \frac{1}{2} \bigg(\frac{1}{t_2^2} + k^2 \bigg ).
    \end{split}
\end{equation}

The covariance between the two noise operators can be calculated as:
\begin{multline}\label{eq:covar}
    \covar(\hat{N}_{x_{out}},\hat{N}_{p_{out}}) = 
     \frac{1-2g^2s^2 + s^4}{1-s^4} \bigg (\frac{1-\eta_H}{\eta_H} \bigg) \frac{k}{2}  \\
     + \frac{2(1-g^2 s^2)(g^2 s^2 - s^4)}{1-s^4}\bigg (\var{(\hat{N}_{p_{out}})} -\var{(\hat{N}_{x_{out}})}   \bigg).
\end{multline}
Note that while the covariance is zero in the cases where phase-shift $\phi = 0$, it attains positive values in the cases where $\phi \neq 0$  with realistic detectors ($\eta_H<1$). Complete derivation of; \eqref{eq:xtel,ptel} - \eqref{eq:covar} can be found in Appendix \ref{sec:appC}. Due to generally non-zero covariance, noise has to be considered carefully. Therefore, the variances and covariances of the noise operator can be formally formulated in the form of a noise matrix $\Sigma$. We define the matrix as follows:
\begin{equation}\label{eq:sigma_matrix}
    \Sigma= \begin{pmatrix}
    \var(\hat{N}_{x_{out}}) & \covar(\hat{N}_{x_{out}},\hat{N}_{p_{out}}) \\
    \covar(\hat{N}_{x_{out}},\hat{N}_{p_{out}}) & \var(\hat{N}_{p_{out}})
    \end{pmatrix}.
\end{equation}

For comparison, we also consider a scenario in which the teleported state is directly squeezed with the help of the measurement-induced scheme \cite{filip_marek}. In this scheme, the input state is mixed with a third squeezed state on a beam splitter with the transmission coefficient $t_0$. The additional mode is then measured, and the obtained value is again used to drive displacement feed-forward operation. The output mode is once again equal to Eq. \eqref{eq:xtel,ptel}. The squeezing parameter $s$ is
\begin{equation}\label{eq:spre}
    s_{BAS} = t_0.
\end{equation}
We call this approach basic squeezing (BAS-sq).
The added noise is the sum of the noise from the basic squeezing circuit and from the quantum teleportation circuit: $ \hat{N}_{x_{out}} = \hat{N}_{x_{bas}} + \hat{N}_{x_{tel}}$, analogically for $\hat{p}_{out}$.
The variances are:
\begin{equation}\label{eq:bas_sq}
\begin{split}
     \var(\hat{N}_{x_{bas}}) & = r_0^2 \big ( \var(\hat{x}_{s_0}) \eta_S + \frac{1}{2}(1-\eta_S ) \big ), \\
     \var(\hat{N}_{p_{bas}}) & = \frac{1}{2} \frac{r_0^2}{t_0^2} \frac{1-\eta_H}{\eta_H}. 
    \end{split}
\end{equation}
The variance of $\var(\hat{N}_{x_{tel}})$ and  $\var(\hat{N}_{p_{tel}})$ are obtained as in Eqs. \eqref{eq:varNx} - \eqref{eq:N1N2} but with $t_1 =t_2 = 1/2$ and $\phi = 0$. The covariance is zero. The $\Sigma$ is then a diagonal matrix.

To see the practical impact of the imperfections in the squeezers, we need to analyze them in the context of specific quantum states. We consider the vacuum state as a stand-in for an arbitrary coherent state that allows testing basic noise properties \cite{furusawa}, and a single photon state that showcases the influence of imperfections on the non-Gaussian nature of the state. Non-Gaussian states with negativity in the Wigner representation are necessary for achieving an advantage in quantum computing \cite{mari}. These effects can be evaluated from the Wigner and characteristic functions of the transformed states. The characteristic function is practical for the evaluation of fidelity \cite{caves2004}, while the Wigner function and its negative regions can be used for recognizing non-Gaussian behavior \cite{mari,kenfack}. The negativity of the Wigner function is only a sufficient indicator of the state’s non-Gaussianity. Its vanishment marks the transition into a classically simulable regime~\cite{mari}, in which the squeezing gate is no longer considered to be of sufficient quality. Combining these approaches helps to avoid unnecessary numerical errors that arise when computing in the Fock basis \cite{provaznik}.
The variance matrix of non-Gaussian states is not sufficient to describe their nature.  

The Wigner function of the transformed state $W_{out} (x,p) $ is equal to the convolution of the squeezed input Wigner $W_{s,\text{in}}(x,p)$ function with added noise $G(x,p)$~\cite{braunstein_loock, braunstein_kimble}:
\begin{equation}\label{eq:convolution}
    W_{out}(x,p) = (W_{s,\text{in}} * G)(x,p).
\end{equation}
The Wigner functions of squeezed vacuum ($\text{in} =0 $) and single photon states ($\text{in} =1 $) are \cite{leonhardt}
\begin{align}
    W_{s,0}(x,p  ) &=\frac{1}{\pi} e^{ -\frac{x^2}{s^2} - s^2p^2 }, \\
    W_{s,1}(x,p) &= \frac{1}{\pi} \bigg(2 \frac{x^2}{s^2} + 2s^2 p^2 -1 \bigg)e^{ -\frac{x^2}{s^2} - s^2p^2 }.
\end{align}
The parameter $s$ is the squeezing parameter defined in Eq. \eqref{eq:xtel,ptel}.
The added noise is Gaussian,
\begin{equation}
    G(x,p) = \frac{1}{\pi \sqrt{\det{\Sigma}}} e ^{-(x,p)\Sigma^{-1} (x,p)^{\text{T}}},
\end{equation}
with a covariance matrix defined in Eq. \eqref{eq:sigma_matrix}.

The characteristic function of the transformed state $\chi_{out}(\xi)$ is equal to the product of the squeezed input characteristic function with added noise $G(\xi)$ \cite{caves2004,dell_anno}
\begin{equation}
    \chi_{out} (\xi) = \chi_{s,\text{in}}(\xi) G(\xi).
\end{equation}

The characteristic functions of squeezed vacuum and single photon states are
\begin{equation}\label{eq:chars}
\begin{split}
    \chi_{s,0} (\xi)&= e^{-1/2 |\alpha|^2} ,\\
    \chi_{s,1} (\xi) &= e^{-1/2 |\alpha|^2} (1- |\alpha|^2),
\end{split}
\end{equation}
where
\begin{equation}
    \alpha = \xi \cosh (\log s) - \xi^* \sinh (\log s).
\end{equation}
The characteristic function of Gaussian noise can be obtained as 
\begin{equation}
    G(\xi)=  e ^{-(\text{Re}\xi,\text{Im}\xi)\Sigma(\text{Re}\xi,\text{Im}\xi)^{\text{T}}}.
\end{equation}
The $\xi$ is a complex number. 

 \begin{figure}[hpbt]
\includegraphics[width=0.5
\textwidth]{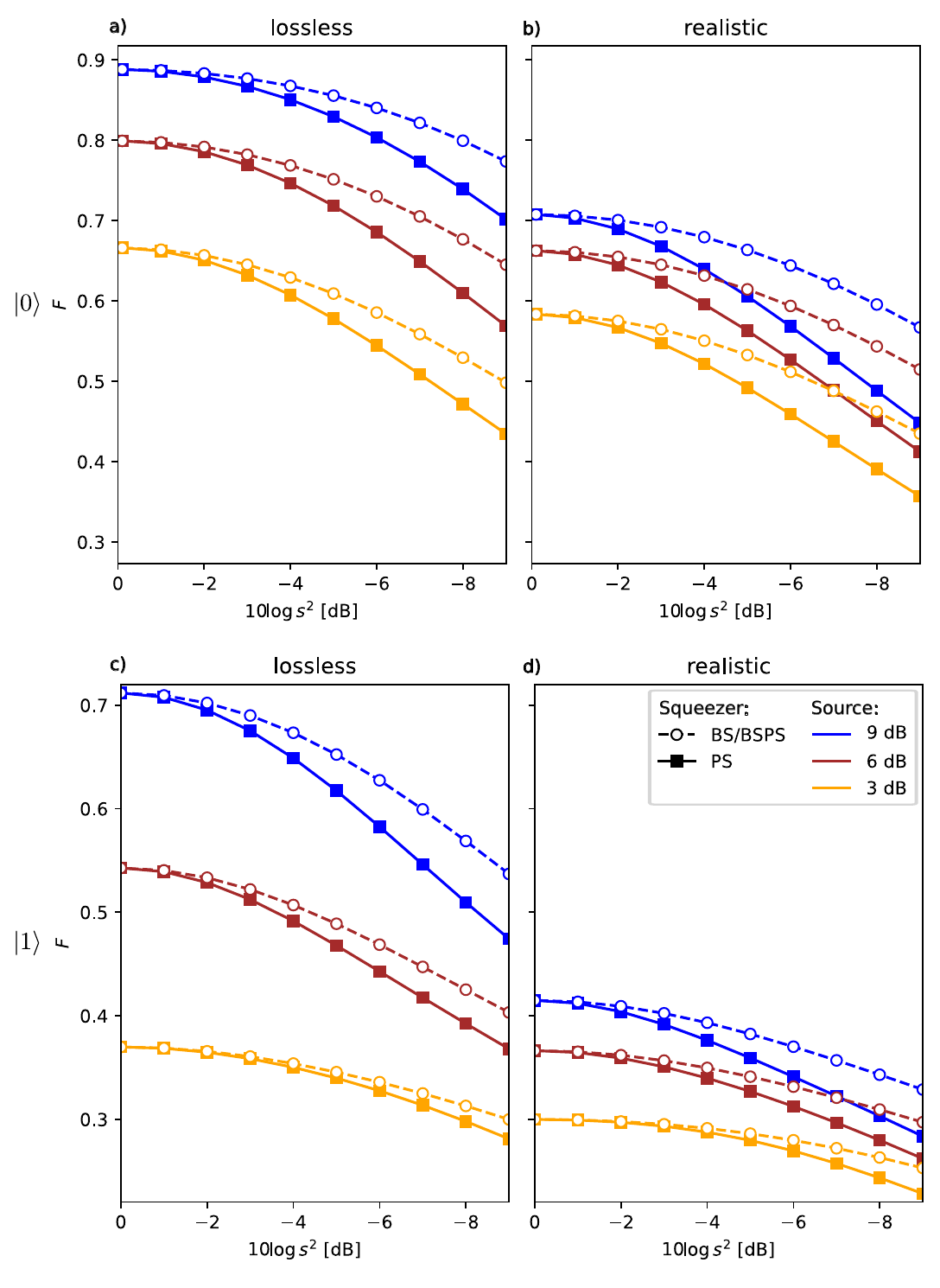}
\caption{\label{fig:fid} The comparison of fidelities for a given squeezing-gate parameter $s$. \textbf{a)} Vacuum state in the lossless scenario; \textbf{b)} vacuum state in the realistic scenario with $\eta_S = 0.8$ and $\eta_H = 0.9$; \textbf{c)} Single photon state in the lossless scenario; \textbf{d)} Single photon state in the realistic scenario with $\eta_S = 0.8$ and $\eta_H = 0.9$. The different lines distinguish different protocols,  PS-sq (full lines with squares) and  BS-sq (dashed lines with circles), and three different squeezing levels of the resource squeezed states, 9 dB (blue), 6 dB (brown), and 3 dB (yellow). The lines for BSPS-sq overlap with the lines for BS-sq. }
\end{figure}

Now, we can define metrics for studying the three squeezers - PS-sq, BS-sq, and BSPS-sq, and for BAS-sq. The used metrics are fidelity \cite{chuang}, minimum of the Wigner function \cite{kenfack}, total noise, entanglement breaking bound, and Genuine non-Gaussianity (GnG) \cite{lachman}. For PS-sq, we directly evaluate the performance. For BS-sq and BSPS-sq, which allow achieving the desired squeezing for whole sets of parameters, we optimize over these sets to obtain maximal fidelity of the squeezed state.  The fidelity is defined as the integral of the product of characteristic functions of the ideally squeezed and transformed quantum states:
\begin{equation}\label{eq:fid_chi}
    F = \frac{1}{\pi}\int G(\xi) |\chi_{s,\text{in}} (\xi)|^2 d^2 \xi,
\end{equation}
where the particular characteristic functions are given by (\ref{eq:chars}). 

The minimum of the Wigner function is relevant only for a single photon state, where it can be used to determine non-Gaussianity. Thanks to the symmetry of the state, the minimum is the value of the Wigner function at the origin, $W(0,0)$. This can be numerically obtained from Eq. \eqref{eq:convolution}.

To analyze the added noise, we have to do two things. We define the total noise as the noise added to the input state before the squeezing. This corresponds to the reparameterization of Eqs. \eqref{eq:xtel,ptel}. We simply factor out the $s$ and $\textstyle \frac{1}{s}$. Therefore: $\hat{N}_{x_{out}} \rightarrow \textstyle \frac{1}{s}\hat{N}_{x_{out}} $ and $\hat{N}_{p_{out}} \rightarrow s \hat{N}_{p_{out}} $. We re-parametrize the noise matrix 
\begin{equation}
    \Sigma' = \Sigma \odot M, \;\;\;\;     M = \begin{pmatrix}
         \frac{1}{s^2} & 1 \\
         1 & s^2
     \end{pmatrix}.
\end{equation}
where symbol $\odot$ denotes element-wise product \cite{hadamard_product}. 
The total noise is defined as the sum of eigenvalues of $\Sigma'$,
\begin{equation}\label{eq:NT}
    N_T = \sum_{i =1}^n \lambda_i, \; \; \; \lambda_i \in \text{spec}(\Sigma').
\end{equation}
The eigenvalues can be used to define a new, diagonal, noise matrix with determinant $N_P$,
\begin{equation}
    N_P = \prod_i^n \lambda_i,
\end{equation}
which can be used to define the entanglement-breaking bound (E-breaking)  
\begin{equation}\label{eq:e-breaking}
    N_P \geq 1.
\end{equation}
This is a sufficient condition for breaking correlations between quadratures when teleporting one mode of the TMSV state. The bound represents, up to Gaussian processing, adding two units of vacuum noise to both quadratures of a single mode, which removes quadrature correlation when applied to one mode of the maximally entangled state.

For the computation of GnG, it is necessary to find the density matrix in a Fock basis. GnG requires knowledge of photo-statistics.
The density matrix of the teleported state is defined as follows:
\begin{equation}
    \hat{\rho}_{out} = \iint G(x,p) \hat{D}\hat{S}\hat{\rho}_{in} \hat{S}^{\dagger}\hat{D}^{\dagger} dx dp,
\end{equation}
where $\hat{S} \equiv \hat{S}(-\log s )$ and $\hat{D} \equiv \hat{D}(x,p)$. The squeezing operator is defined as \cite{weedbrook,leonhardt}: 
\begin{equation}\label{eq:squeezing}
    \hat{S}(r) = e^{\frac{i}{2} r (\hat{x}\hat {p} - \hat{p}\hat{x})}. 
\end{equation}
The parameter $r$ is a general parameter. The displacement operator is defined as \cite{weedbrook,leonhardt}: 
\begin{equation}\label{eq:displacement}
    \hat{D}(x_0,p_0) = e^{i(p_0\hat{x} - q_0 \hat{p} )}.
\end{equation}

In the next section, we analyze PS, BS, and BSPS-sq to find the superior teleportation-based squeezer in CV teleportation.

\begin{figure}[hpbt]
\includegraphics[width=0.5
\textwidth]{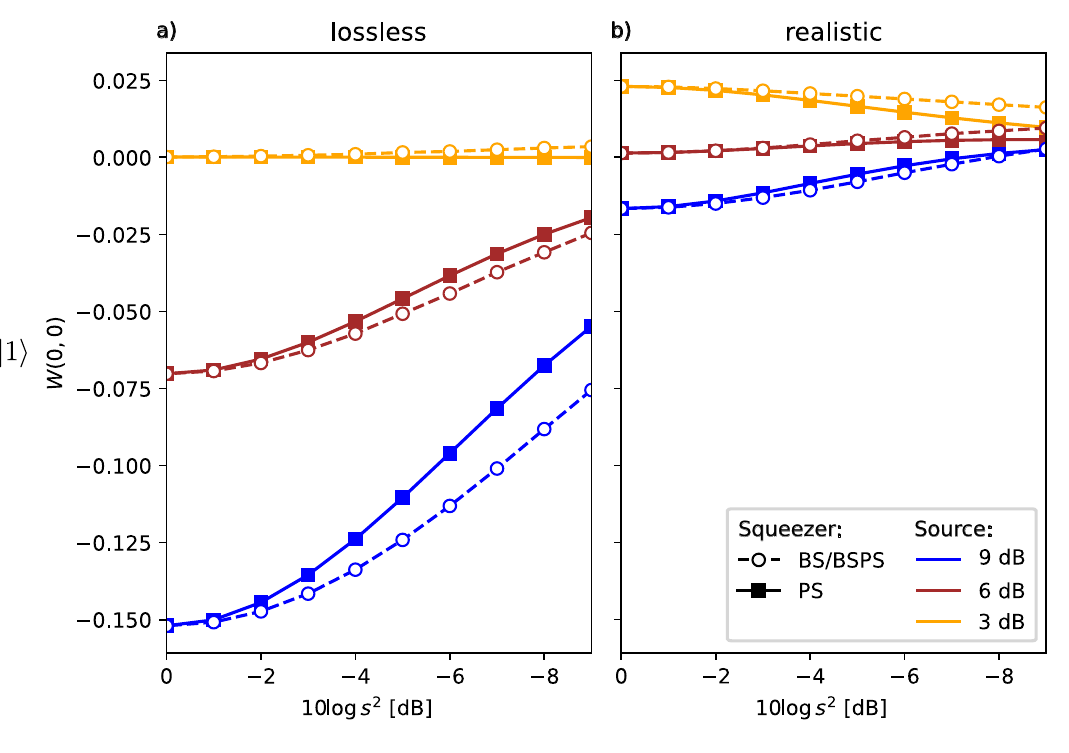}
\caption{\label{fig:negativity}  The comparison of values of the $W(0,0)$ for the given target squeezing $s$ of the single photon state for teleportation-based squeezer. \textbf{a)} Lossless scenario; \textbf{b)} realistic scenario with $\eta_S = 0.8$ and $\eta_H = 0.9$. The different lines distinguish different squeezers,  PS-sq (full lines with squares) and BS-sq (dashed lines with circles), and three different squeezing levels of the resource squeezed states, 9 dB (blue), 6 dB (brown), and 3 dB (yellow). The lines for BSPS-sq overlap with the lines for BS-sq. }
\end{figure}

\section{Results}\label{sec:analysis}
In the following, we shall compare the three teleportation-based squeezers, PS-sq, BS-sq, and BSPS-sq.  We will employ teleportation-based squeezers to squeeze two kinds of input states, the vacuum state and the single photon state, and compare the results for various levels of parameter squeezing $s$ and available resources. Here, the squeezing parameter $s$ stands for the individual squeezing parameters  $s_{BS}$, $s_{PS}$, and $s_{BSPS}$, and its value will be the same for all the squeezers. In the case of BS-sq and BSPS-sq, in which the squeezing parameter can be obtained by a set of optimization parameters $t_1$, $t_2$, and $\phi$, those will be optimized for the best performance given by fidelity with the ideally squeezed state, with no noise terms. The numerically obtained optimal parameters can be found in Appendix \ref{sec:appD}. We plot the fidelity, the Wigner negativity, total added noise, and product of variances. The optimization with respect to total noise can be found in the supplement material \ref{sec:appE}. There, we compare the two fidelities and the total noise. The different resources consist of squeezed vacuum states with variances of the squeezed quadrature reduced by 3, 6, and 9 dB below the level of vacuum fluctuations. We will consider the same resource squeezing in all relevant quadrature operators $\hat{x}_{s1}$ and $\hat{p}_{s2}$. We also compare the lossless scenario, in which the resource states are pure and detectors have unit quantum efficiency, to realistic scenarios, in which the resource squeezed states undergo losses with $\eta_S = 0.8$ and the detectors have limited quantum efficiency $\eta_H = 0.9$. All functions depend on $10\log_{10}(s^2)$. The results for BAS-sq can be found in Appendix \ref{sec:appF}. We can already see from Eq. \eqref{eq:bas_sq} that this approach introduces even more noise in the form of noise operators $\hat{N}_{x_{bas}}$ and $\hat{N}_{p_{bas}}$ and thus is not relevant in the current analysis, but it gives us hints what should we expect when we add another homodyne detector and non-ideal source of squeezing.

The obtainable fidelities for the considered scenarios are shown in Fig.~\ref{fig:fid}.  We can see that, even though the specific values differ, the qualitative behavior is the same across all the scenarios. The quality of the operation depends on the squeezing parameter $s$ and diminishes as it increases.  The quality of the operation strongly depends on the resource squeezing, but we can see that BS-sq consistently outperforms PS-sq and matches the BSPS-sq method. In the realistic case, we can see that BS-sq with lower resource squeezing outperforms PS-sq with higher resource squeezing. This concerns the case where $10 \log s^2 = -7$ dB and PS-sq with 9 dB of resource squeezing and BS-sq with 6 dB of resource squeezing. 

To demonstrate how the squeezing operation affects negativity of the Wigner function of the single photon case, Fig. \ref{fig:negativity} shows the values of the Wigner functions at the origin, $W(0,0)$. We stress that we used optimization of fidelity, and we plot minima of states given by the optimized parameters computed in the previous step. We can again see that the qualitative behavior is consistent with the behavior of fidelities in Fig.~\ref{fig:fid}. The BS-sq scheme outperforms the PS-sq scheme in all relevant scenarios ($W(0,0) <0$), even though the differences begin to smear in the realistic scenario. We can also see that 3 dB of resource squeezing in the lossless case and 6 dB of resource squeezing in the realistic case are insufficient to preserve the state's negativity. The minimum of Wigner negativity could be subject to optimization instead of fidelity, leading to sub-optimal fidelity. We could obtain the output states with the same or lower negativity as in the case of fidelity optimization, but we would not have guaranteed similarity of the output states with the squeezed input states, specifically for the low source squeezing, where the negativity is suppressed with added noise. Therefore, the fidelity of the output state with the squeezed input state would go to zero. 
\begin{figure}[hpbt]
\includegraphics[width=0.5
\textwidth]{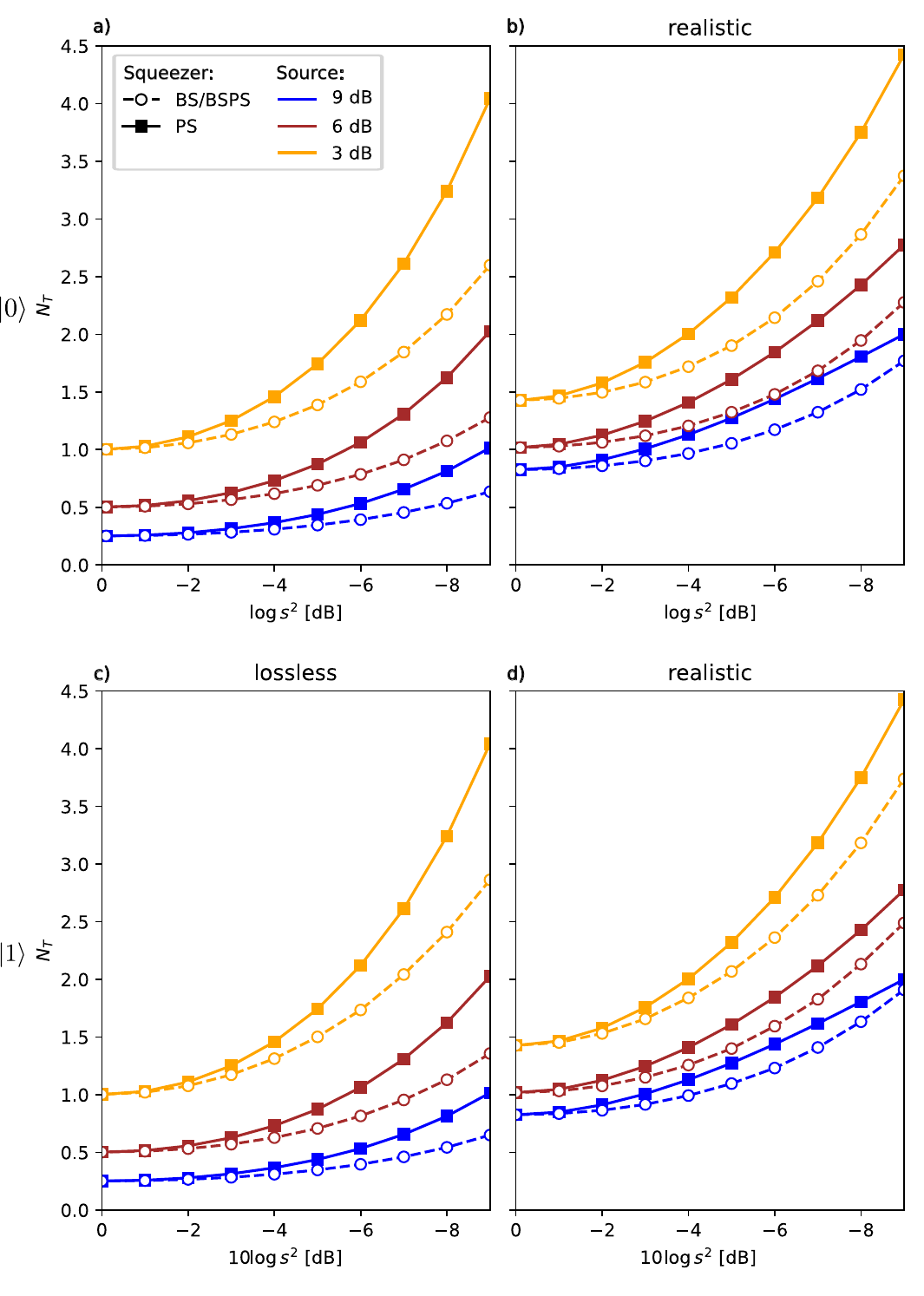}
\caption{\label{fig:total_noise}
The comparison of total noise $N_T$ (noise added before squeezing) for a given squeezing parameter $s$. \textbf{a)} Vacuum state in the lossless scenario; \textbf{b)} vacuum state in the realistic scenario with $\eta_S = 0.8$ and $\eta_H = 0.9$; \textbf{c)} Single photon state in the lossless scenario; \textbf{d)} Single photon state in the realistic scenario with $\eta_S = 0.8$ and $\eta_H = 0.9$. The different lines distinguish different protocols,  PS-sq (full lines with squares) and  BS-sq (dashed lines with circles), and three different squeezing levels of the resource squeezed states, 9 dB (blue), 6 dB (brown), and 3 dB (yellow). The lines for BSPS-sq overlap with the lines for BS-sq.}
\end{figure}
\begin{figure}[hpbt]
\includegraphics[width=0.5
\textwidth]{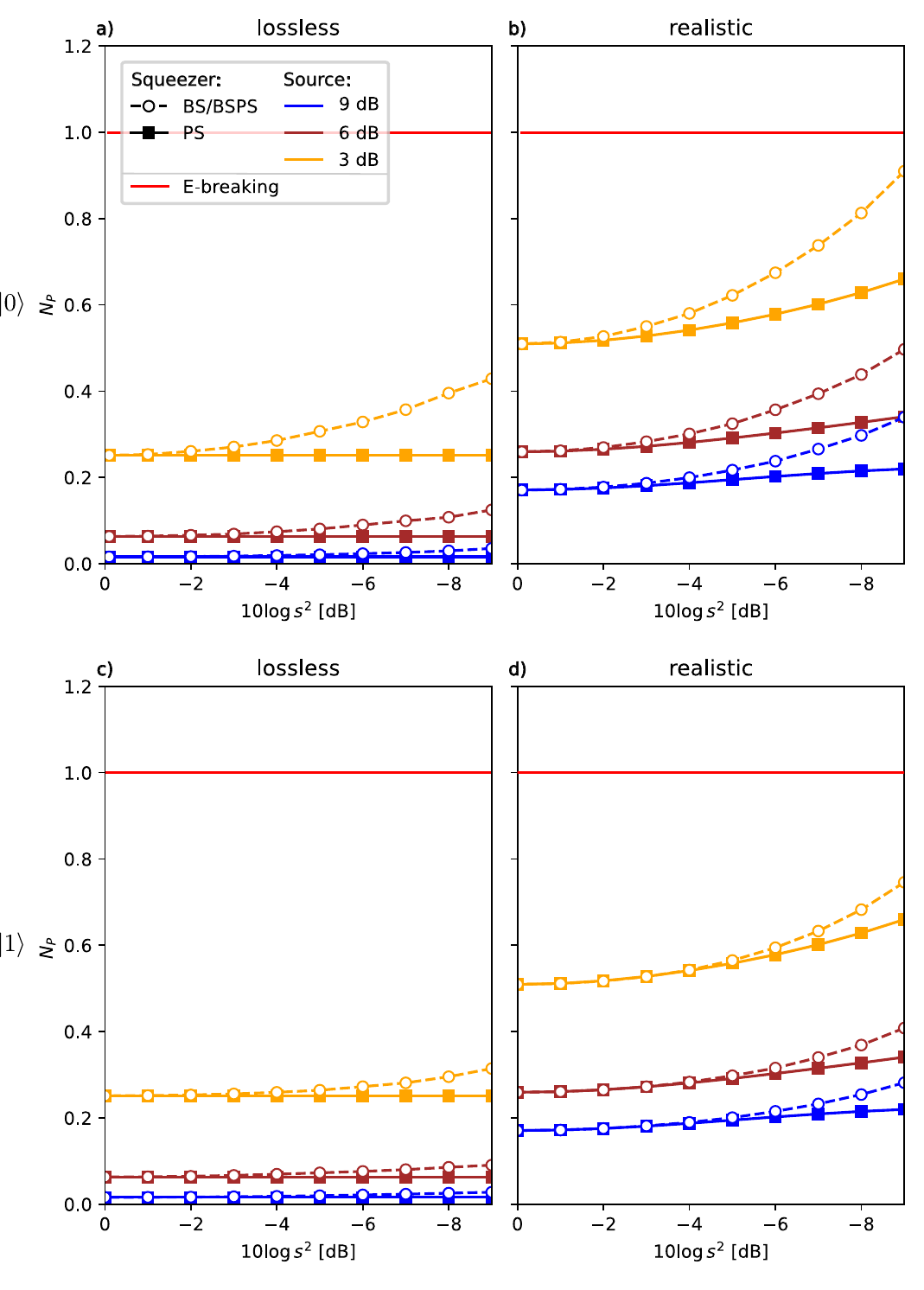}
\caption{\label{fig:product_noise}
The comparison of the product of variances of noise operators $N_P$ for a given squeezing parameter $s$.\textbf{a)} Vacuum state in the lossless scenario; \textbf{b)} vacuum state in the realistic scenario with $\eta_S = 0.8$ and $\eta_H = 0.9$; \textbf{c)} Single photon state in the lossless scenario; \textbf{d)} Single photon state in the realistic scenario with $\eta_S = 0.8$ and $\eta_H = 0.9$. The different lines distinguish different protocols,  PS-sq (full lines with squares) and BS-sq (dashed lines with circles), and three different squeezing levels of the resource squeezed states, 9 dB (blue), 6 dB (brown), and 3 dB (yellow). The lines for BSPS-sq overlap with the lines for BS-sq. 
The value of $1$ corresponds to the entanglement breaking bound, which is denoted by the red full line.}
\end{figure}

The behavior of both fidelity, Fig.~\ref{fig:fid}, and Wigner function negativity, Fig.~\ref{fig:negativity}, is related to the behavior of total noise Eq. \eqref{eq:NT}, shown in Fig.~\ref{fig:total_noise}. After the correction for the squeezing $s$, we can see that the total noise added during the protocol monotonically increases with the squeezing parameter $s$. The correction for the squeezing is a necessary step. Without it, the total added noise is lower for the PS-sq, which adds the same noise into both quadratures, see Appendix \ref{sec:appE}. After the correction, however, the initially asymmetrical noise of BS-sq gets compensated and ends up being lower. Note that the noise added during the squeezing depends on the initial state. This is because the parameters were optimized for maximal fidelity.  
In fact, the total noise provides a partial explanation of the behavior of fidelity in Fig.~\ref{fig:fid}. We have found a counterexample where the rule: the less the total noise, the higher the fidelity, does not hold. The counterexample was obtained by the minimization of total added noise. We also performed the optimization with respect to added noise. We obtained qualitatively similar results as in the case of optimization to fidelity when considering fidelity, see Fig. \ref{fig:Deltas} in Appendix \ref{sec:appE}.

The entanglement breaking bound is easily visible in Fig.~\ref{fig:product_noise}, it is the red line because Fig.~\ref{fig:product_noise} shows the product of variances. Every studied case is under the threshold, but the realistic case of PS squeezer approaches the threshold for target squeezing $10 \log s^2 \approx -9$ dB. This indicates that the distribution of noise in the PS-sq case is more even in both quadratures than in the case of BS-sq.

Finally, we can also evaluate the non-Gaussianity of the squeezed single photon state with the help of the first order of GnG.  Interestingly, while BS-sq offers higher fidelities for given $s$, see Fig.~\ref{fig:fid}, it also leads to quicker loss of non-Gaussianity certified by GnG, see Fig. \ref{fig:gng_supp} in Appendix \ref{sec:appG}. This implies that the squeezed states produced by BS-sq contain higher contributions of the vacuum state in their photo-statistics. The BS-squeezer has no advantage when GnG is used for the non-Gaussian certification of the teleport-squeezed single photon states.

The poor performance of BAS-sq is caused by the additional resource of squeezing and an additional homodyne detector, both of which contribute to extra noise in the output state. See Appendix \ref{sec:appF}, where we compare BAS-sq with PS-sq in fidelity and in the minimum of the Wigner function. 

From a theoretical point of view, the squeezing parameter $s$ defined in Eqs. \eqref{eq:squeezing_id} and \eqref{eq:xtel,ptel} can be relaxed: $s \rightarrow s + \Delta s$ in one quadrature and left unchanged in the second one. This relaxation would lead to optimization of fidelity between the mixed state and the teleported squeezed state, which is an ill-defined procedure. The relaxation of the squeezing parameter $s$ remains open and can be considered with the close connection to the potential experiment. 

The optimization with respect to added noise was performed and led to qualitatively the same results as in the case of maximization of fidelity, see Appendix \ref{sec:appE}. The conclusion holds for the case of maximization of fidelity and for the case of minimization of total noise. The optimized added noise can be used as a baseline for evaluating non-Gaussian features of other non-Gaussian states \cite{petrmarek,simon}. The optimization with respect to the minimum of the Wigner function is not universal in the sense of Gaussian and non-Gaussian input states. The optimization does not work, especially for low-resource squeezing, namely, the optimization finds such parameters of the BS squeezer that the resulting state has low fidelity with the target squeezed state. 
The next step should aim for the experimental realization of the squeezing gate in a cluster state using the experimentally obtainable unbalanced beam splitters \cite{cat,cat2}. Hand in hand with the experiment, the analysis can be extended to relax the strict conditions on amplitude transmission coefficients: $t_1 \rightarrow t_1 + \Delta_{t}$ and $t_2 \rightarrow t_2 + \Delta_{t}$. The analysis of the multi-teleportation channel can also be considered, but it is expected that the accumulated noise will suppress every quantum feature of the input state \cite{verma}.

\section{Conclusion}\label{sec:conclusion}
We compared different squeezing gates that can be used in bosonic cluster state computation. The squeezer based on variable beam splitters, BS-sq, outperforms the squeezer based on variable phase shift, PS-sq, in ideal and realistic scenarios, and it matches the fully general BSPS-sq scheme in performance. We used vacuum and single photon states as specific examples of Gaussian and non-Gaussian states, respectively. BS-squeezer outperforms the PS-squeezer, regarding both fidelity and negativity in the Wigner function of the teleport-squeezed state. In both fidelity and negativity, the advantage of BS-sq can be seen for a high level of target squeezing given by parameter $10\log s^2 \approx  -9$ dB. For this regime, the case of the vacuum state, BS-sq gains around 16\% of fidelity performance in a realistic scenario.  In the case of a single photon state, the resource squeezing plays a significant role. The fidelity is generally lower for the same resource squeezing compared to the vacuum state. However, the gain of the fidelity performance of BS-sq is up to 8\% for resource squeezing of 9 dB and up to 2\% for 3 dB of resource squeezing under realistic conditions. The single photon is more fragile to losses; therefore, for low resource squeezing, the negativity of the Wigner function is lost during the teleportation for every teleportation-based squeezer. If the resource squeezing is $>6$ dB, the negativity can survive for a small amount of teleportation-based squeezing; the differences between PS-sq and BS-sq are negligible.
The advantage of BS-sq in terms of fidelity and negativity performance lies in the broader parameter space compared to PS-sq. 

PS-sq requires only the application of three phase shift operations, rather than two variable beam splitters of the BS-sq protocol. This simplification is paid for by worse performance. Fortunately, since both protocols merge in the limit of infinite resources, the deficiency of the protocol can be compensated for by using more strongly squeezed resources. Still, for near-term experimental tests, especially in scenarios where the squeezing does not need to be controlled dynamically, squeezing by BS-sq is bound to provide better performance. 


\
\section*{ACKNOWLEDGMENTS}
P.M. and M.M. acknowledge support of the Czech Science Foundation (project 25-17472S). We acknowledge the European Union's HORIZON Research and Innovation Actions under Grant Agreement no. 101080173 (CLUSTEC) and project CZ.02.01.010022$\_$0080004649 (QUEENTEC) of the EU and the Czech Ministry of Education, Youth and Sport (MEYS). M.M. acknowledges IGA-PrF-2025-010. R.F. and M.M. further acknowledge the project 8C24003, the MEYS CR, and EU under Grant Agreements no. 101017733 and 731473 (CLUSSTAR).
M.M. acknowledges discussions with Jan Provazník, Šimon Br\"aue,r and Vojtěch Kala. 

\section*{Data availability}
The datasets supporting the presented results are publicly available \cite{data}.
\bibliography{biblio}

\appendix

\section{Quadrature transformations}\label{sec:appA}
We introduce Gaussian unitaries, which were used in the derivation of relations between input and output quadratures \cite{braunstein_loock,weedbrook,leonhardt}.
The beam-splitter unitary is defined
\begin{equation}
    \hat{B}(\theta) = e^{i \Theta( \hat{q}_1 \hat{p}_2- \hat{p}_1 \hat{q}_2 )}
\end{equation}
A beam splitter transformation transforms quadratures as follows:
$\hat{x}_1 \rightarrow t \hat{x}_2  + r \hat{x}_1  $, the same holds for $ \hat{p}_1 $ and $\hat{x}_2 \rightarrow t\hat{x}_1  - r\hat{x}_2  $ analogously for $\hat{p}_2$. The parameter $t^2 = \cos^2 (\Theta) $ is called transmissivity, reflectivity is defined as $r^2 = \sin^2 (\Theta) $.

The phase rotation operator is defined as 
\begin{equation}
    \hat{R}(\phi) = e^{i \phi \hat{n},}
\end{equation}
where $\hat{n} = \hat{a}^{\dagger} \hat{a}$ is a number operator. The quadratures are rotated: $\hat{x} \rightarrow \hat{x} \cos \phi - \hat{p} \sin{\phi}$ and $\hat{p} \rightarrow \hat{p} \cos \phi + \hat{x} \sin{\phi}$. 

We define squeezing operator $\hat{S}(r)$ as 
\begin{equation}\label{eq:squeezing}
    \hat{S}(r) = e^{\frac{i}{2} r (\hat{x}\hat {p} - \hat{p}\hat{x})}. 
\end{equation}
If we substitute the general variable $r$ with the parameter $s$ in the following way: $r = -\ln{s}$, the squeezing operator scales the quadratures: $\hat{x} \rightarrow \textstyle g \hat{x}$, $\hat{p} \rightarrow \frac{1}{g} \hat{p}$.  

The quadratic phase operator $\hat{P}(k)$ is defined  as follows 
\begin{equation}\label{eq:shearing}
    \hat{P} (k) = e^{\frac{i}{2} k \hat{x}^2}.
\end{equation}
The shearing operator transforms the quadratures in the following way: $\hat{x} \rightarrow \hat{x}$ and $\hat{p} \rightarrow \hat{p} + k\hat{x} $. 

The displacement operator is defined as 
\begin{equation}
    \hat{D}(x_0,p_0) = e^{i(p_0 \hat{x} - q_0 \hat{p} )}.
\end{equation}

\section{Decomposition of shear gate and product of shear and squeezing}\label{sec:appB}
The shear gate in Eq. \eqref{eq:shearing} can be decomposed as follows:
\begin{equation}\label{eq:decomposition_P}
    \hat{P}(k) = \hat{R}(\zeta) \hat{S}(\xi) \hat{R}(\zeta - \textstyle \frac{\pi}{2}),
\end{equation}
where 
\begin{align}
    e^{-2\xi} &= \frac{1}{2} \bigg (2+k^2-k\sqrt{4+k^2} \bigg ), \\ \cos{(\alpha)} & =  \frac{1}{\sqrt{e^{-2\xi} + 1 }}.
\end{align}
The product of squeezing and shear gates can be similarly decomposed as 
\begin{align}\label{eq:decompositon_SP}
\hat{S}(r) \hat{P}(k) = \hat{R}(\zeta) \hat{S} (\xi) \hat{R} (\epsilon),
\end{align}
where 
\begin{align}
\begin{split}
    e^{-2\xi} &= \frac{1}{2 e^{-2r}}\bigg(1+ e^{-4r}+k^2\\ &- \sqrt{(1-e^{-4r})^2 + 2 (1+e^{-4r})k^2 + k^4 }\bigg), \\ \label{eq:alpha,beta}
    \end{split}
\end{align}
and 
\begin{equation}
        \cos(\zeta) = \sqrt{\frac{e^{2\xi} -e^{-2r}} {e^{2 \xi} -e^{-2 \xi}}}, \; \; \cos(   \epsilon) = -\sqrt{\frac{e^{-2\xi} - e^{2r}}{e^{-2\xi} - e^{2 \xi}}}.
\end{equation}
For $e^{-2r} =1 $ cosine of the angel $\epsilon$ is  $-\cos{(\epsilon)} = \sin(\zeta)$, the angel itself is  $\epsilon = \zeta -\textstyle \frac{\pi}{2}$. Therefore, we reproduce the decomposition \eqref{eq:decomposition_P}.

\section{Details on teleportation-based squeezer}\label{sec:appC}
We start with the simplified scheme in Fig. 1 in the main text. The components \text{PS}($-\zeta$), and \text{PS}($-\epsilon$) are ignored as well as losses represented by \text{BS}($\sqrt{\eta_S}$). We derive relations between input and output quadratures. Due to the mentioned simplifications, we denoted the output quadratures with a zeta index $\hat{x}^{(\zeta)}_{out}$ and $\hat{p}^{(\zeta)}_{out}$.
The detectors HD$_\text{A}$ and HD$_\text{B}$ measure 
\begin{equation}
    \begin{split}
    \hat{q}_A &\equiv \sqrt{\eta_H} \hat{x}_A + \sqrt{1 - \eta_H} \hat{x}_{\tilde{A}}^0 ,\\ 
    \hat{q}_B &\equiv( \sqrt{\eta_H} \hat{x}_B + \sqrt{1 - \eta_H} \hat{x}_{\tilde{B}}^0) \sin{(\phi)}\\ &+  (\sqrt{\eta_H} \hat{p}_B + \sqrt{1 - \eta_H} \hat{p}_{\tilde{B}}^0 )\cos{(\phi)},
    \end{split}
\end{equation}
where $\hat{q}_A$ are $\hat{q}_B$ the measured quadrature. Due to phase shift, \text{PS}($\phi$), the detector HD$_\text{B}$ measures rotated quadrature by the angle $\phi$.
The output quadratures are
\begin{equation}
\begin{split}
    \hat{x}'_{out} &= t_1 \hat{x}_{s_1} - r_1 \hat{x}_{s_2} + j_1 \hat{q}_A, \\
    \hat{p}'_{out} & = t_1 \hat{p}_{s_1} - r_1 \hat{p}_{s_2} + j_2 \hat{q}_B + j_3 \hat{q}_A, 
    \end{split}
\end{equation}
where $j_i$, $i \in {1,2,3}$ are teleportation gains. 
If $j_i$ are set as follows
\begin{align}\label{eq:gains}
    j_1 = \frac{r_1}{t_1 t_2}, \;\;\;\;
    j_2 = \frac{t_1}{r_1 r_2}\frac{1}{ \cos{\phi}}, \;\;\;\;
    j_3 = \frac{t_1}{r_1 t_2} \tan{\phi}.
\end{align}
We obtain a unity-gain generalized teleportation scheme. 
After the substitution, we get the output quadratures 
\begin{equation}
\label{eq:xtel,ptel_appendix}
\begin{split}
    \hat{x}^{(\zeta)}_{out} &= g \hat{x}_{in} + \hat{N}^{(\zeta)}_{x_{out}}, \\
    \hat{p}^{(\zeta)}_{out} &= \frac{1}{g} \hat{p}_{in} +\frac{k}{g } \hat{x}_{in} +  \hat{N}^{(\zeta)}_{p_{out}},
\end{split}
\end{equation}
where $g$ and $k$ are
\begin{align}\label{eq:g and k}
    g = \frac{r_1 r_2}{t_1 t_2}, \; \; \;
    k = \frac{1}{t_2^2} \tan{\phi}. 
\end{align}
The parts of transformations that are $\propto \hat{x}_{in}$ and $\propto \hat{p}_{in}$ are obtained by shearing and squeezing the input state quadratures, cf. Eqs. \eqref{eq:squeezing} and \eqref{eq:shearing}. 
The noise term $\hat{N}^{(\zeta)}_{x_{out}}$ is composed of two parts: the first part is $\propto \hat{x}_{s_1}$ and is a consequence of the finite source squeezing. We include the losses which are simulated by \text{BS}($\sqrt{\eta_S}$) by the reparameterization $\hat{x}_{s_1} \rightarrow \sqrt{\eta_S}\hat{x}_{s_1} + \sqrt{1-\eta_S} \hat{x}_{\tilde{s_1}}^0$. The mode denoted by the tilde ${\tilde{s_1}}$ (or ${\tilde{s_2}}$) is the auxiliary vacuum mode.  We also introduce an imperfect homodyne detector; the efficiency of such a detector is again simulated by an added beam splitter \text{BS}($\sqrt{\eta_H}$). The same holds for $\hat{N}^{(\zeta)}_{p_{out}}$. Specifically: 
\begin{equation}\label{eq:N'}
\begin{split}
    \hat{N}^{(\zeta)}_{x_{out}} &= \frac{1}{t_1}  \big( \sqrt{\eta_S}\hat{x}_{s_1} + \sqrt{1-\eta_S} \hat{x}_{\tilde{s_1}}^0\big)\\  &+ \sqrt{\frac{1-\eta_H}{\eta_H}} \frac{g}{r_2} \hat{x}^0_{\tilde{A}}, \\
    \hat{N}^{(\zeta)}_{p_{out}} &=  - \frac{1}{r_1} \big( \sqrt{\eta_S}\hat{p}_{s_2} + \sqrt{1 - \eta_S} \hat{p}_{\tilde{s_2}}^0\big) \\ &+\sqrt{\frac{1-\eta_H}{\eta_H}} \frac{1}{g}\bigg \{  \frac{1}{ t_2} \hat{p}^0_{\tilde{B}} +  k \bigg({ t_2} \hat{x}^0_{\tilde{B}}   + {r_2}\hat{x}^0_{\tilde{A}} \bigg) \bigg \}.
\end{split}
\end{equation}
As above, the modes with tilde  $\tilde{A}$ (or $\tilde{B}$) are auxiliary vacuum modes coming into the \text{BS}($\sqrt\eta_H$).
\begin{figure}[hpbt]
\includegraphics[width=0.5
\textwidth]{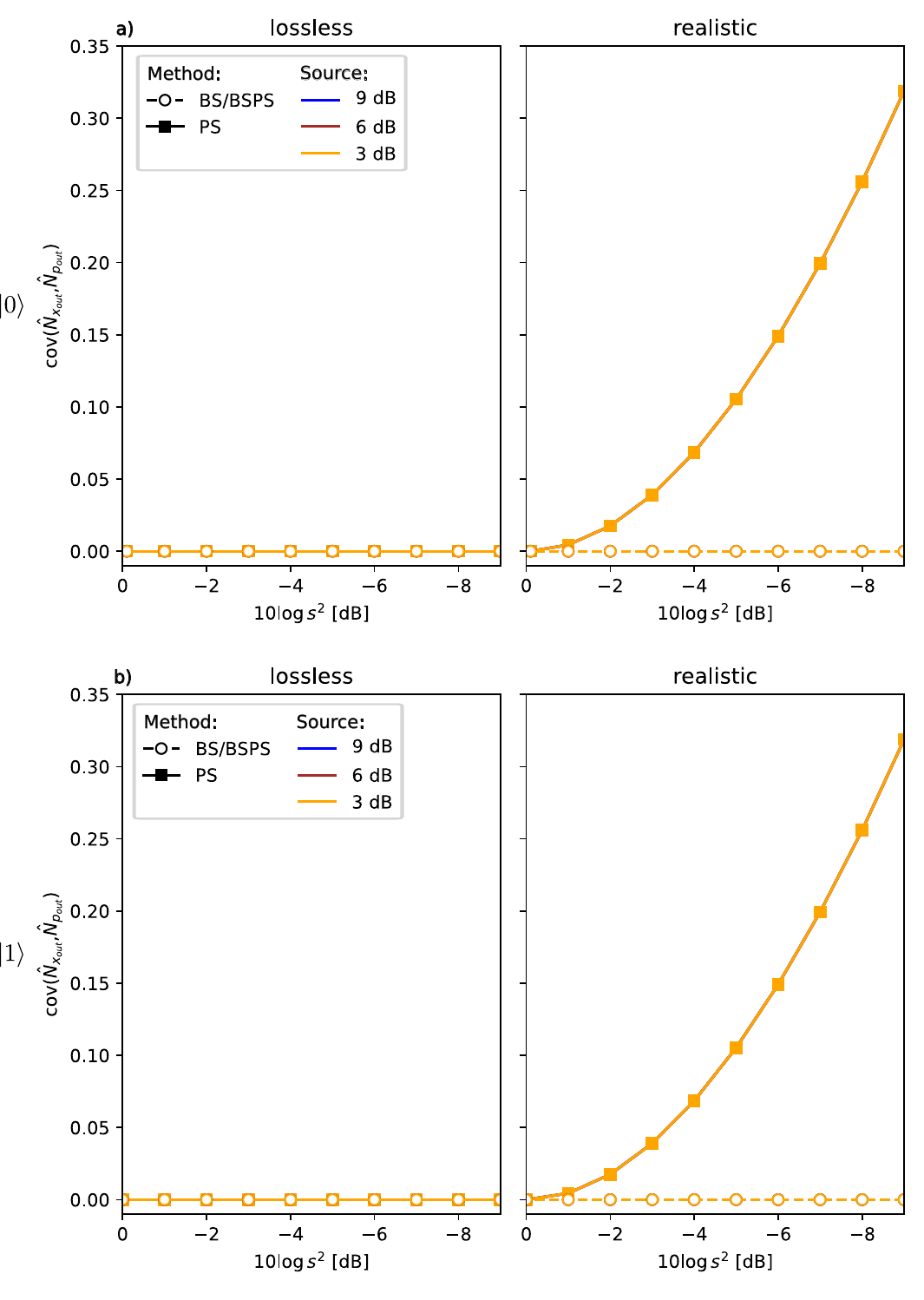}
\caption{\label{fig:cov} The covariance of $\covar(\hat{N}_{x_{out}},\hat{N}_{p_{out}})$.  \textbf{a)} shows on the left lossless case of vacuum state, $\eta_S =1$, $\eta_H = 1$, where is the covariance zero for all the cases. On the right there is a realistic case $\eta_S =0.8$, $\eta_H = 0.9$, it is not zero for the case $k\neq 0$ and $\eta_H < 1$. \textbf{b)} shows the same, but for the case of a single photon state. The different lines distinguish different protocols, PS-sq (full lines with squares) and BS-sq (dashed lines with circles), and three different squeezing levels of the resource squeezed states, 9 dB (blue), 6 dB (brown), and 3 dB (yellow). The lines for BSPS-sq overlap with the lines for BS-sq. } 
\end{figure}

The quadrature transformation in Eq. \eqref{eq:xtel,ptel_appendix} is equivalent to a series of $\hat{S}(-\log g)$ and $\hat{P}(k)$ acting on input quadratures plus noise. This squeezing and shear gate sequence can be decomposed as a series of rotations and squeezing, see Eq.~\eqref{eq:decompositon_SP}. We want to compensate for any rotation in the output quadrature to get axis-aligned squeezing. Therefore, we introduce auxiliary phase rotations \text{PS}($-\zeta$) and \text{PS}($-\epsilon$). These rotations compensate the rotation operators in Eq.~\eqref{eq:decompositon_SP} and we get the final input-output relations:
\begin{equation} \label{eq:xtel,ptel_appendix_rotated}
    \begin{split}
    \hat{x}_{out} &= s \hat{x}_{in} + \hat{N}_{x_{out}}, \\
    \hat{p}_{out} &= \frac{1}{s} \hat{p}_{in}  +  \hat{N}_{p_{out}},
    \end{split}
\end{equation}
where
\begin{equation} 
\begin{split}
 s &= \frac{1}{g\sqrt{2}}\bigg(1+ g^4+k^2 \\ &- \sqrt{(1-g^4)^2 + 2 (1+g^4)k^2 + k^4} \bigg)^{1/2}.
 \end{split}
\end{equation}
The noise operators are rotated by the $-\zeta$:
\begin{equation}
    \begin{split}
        \hat{N}_{x_{out}} &= \hat{N}^{(\zeta)}_{x_{out}} \cos{(-\zeta)} -  \hat{N}^{(\zeta)}_{p_{out}}\sin{(-\zeta)}, \\
        \hat{N}_{p_{out}} &= \hat{N}^{(\zeta)}_{p_{out}} \cos{(-\zeta)} + \hat{N}^{(\zeta)}_{x_{out}}\sin{(-\zeta)}. \\
    \end{split}
\end{equation}
If we substitute the expression for $\cos{(\zeta)}$ from Eq.~\eqref{eq:alpha,beta} and find the variance, we reproduce  Eqs.~(8)-(9) in the main text:
\begin{equation}\label{eq:var_supp}
\begin{split}
\var (\hat{N}_{x_{out}}) &= \frac{(1-g^2 s^2)}{1-s^4} \var{(\hat{N}^{(\zeta)}_{x_{out}})}\\ &+ \frac{(g^2 s^2- s^4)}{1-s^4} \var{(\hat{N}^{(\zeta)}_{p_{out}})}, \\
\var (\hat {N}_{p_{out}}) &= \frac{(1-g^2 s^2)}{1-s^4} \var{(\hat{N}^{(\zeta)}_{p_{out}})} \\ &+ \frac{(g^2 s^2- s^4)}{1-s^4} \var{(\hat{N}^{(\zeta)}_{x_{out}})},
\end{split}
\end{equation}
where the rotated variances are
\begin{equation}
    \begin{split}
        \var{(\hat{N}^{(\zeta)}_{x_{out}})} & = \frac{1}{t_1^2} \big ( \var(\hat{x}_{s_1}) \eta_S + \frac{1}{2}(1-\eta_S ) \big ) \\ &+  \frac{1-\eta_H}{\eta_H} \frac{g^2}{r_2^2} \frac{1}{2}, \\
        \var{(\hat{N}^{(\zeta)}_{p_{out}})} & = \frac{1}{r_1^2} \big ( \var(   \hat{p}_{s_2}) \eta_S + \frac{1}{2}(1-\eta_S ) \big ) \\ &+  \frac{1-\eta_H}{\eta_H} \frac{1}{g^2} \frac{1}{2} \bigg(\frac{1}{t_2^2} + k^2 \bigg ).
    \end{split}
\end{equation}
The formula for the covariance $\covar(\hat{N}_{x_{out}},\hat{N}_{p_{out}})$ of noise operators can be obtain in similar fashion:
\begin{equation}
    \begin{split}
        \covar(\hat{N}_{x_{out}},\hat{N}_{p_{out}}) &= 
     \frac{1-2g^2s^2 + s^4}{1-s^4} \bigg (\frac{1-\eta_H}{\eta_H} \bigg) \frac{k}{2} 
     \\ &+ \frac{2(1-g^2 s^2)(g^2 s^2 - s^4)}{1-s^4} \\ & \times \bigg(\var{(\hat{N}_{p_{out}})} -\var{(\hat{N}_{x_{out}})}   \bigg).
    \end{split}
\end{equation}
\begin{figure*}[hpbt]
\includegraphics[width=0.9
\textwidth]{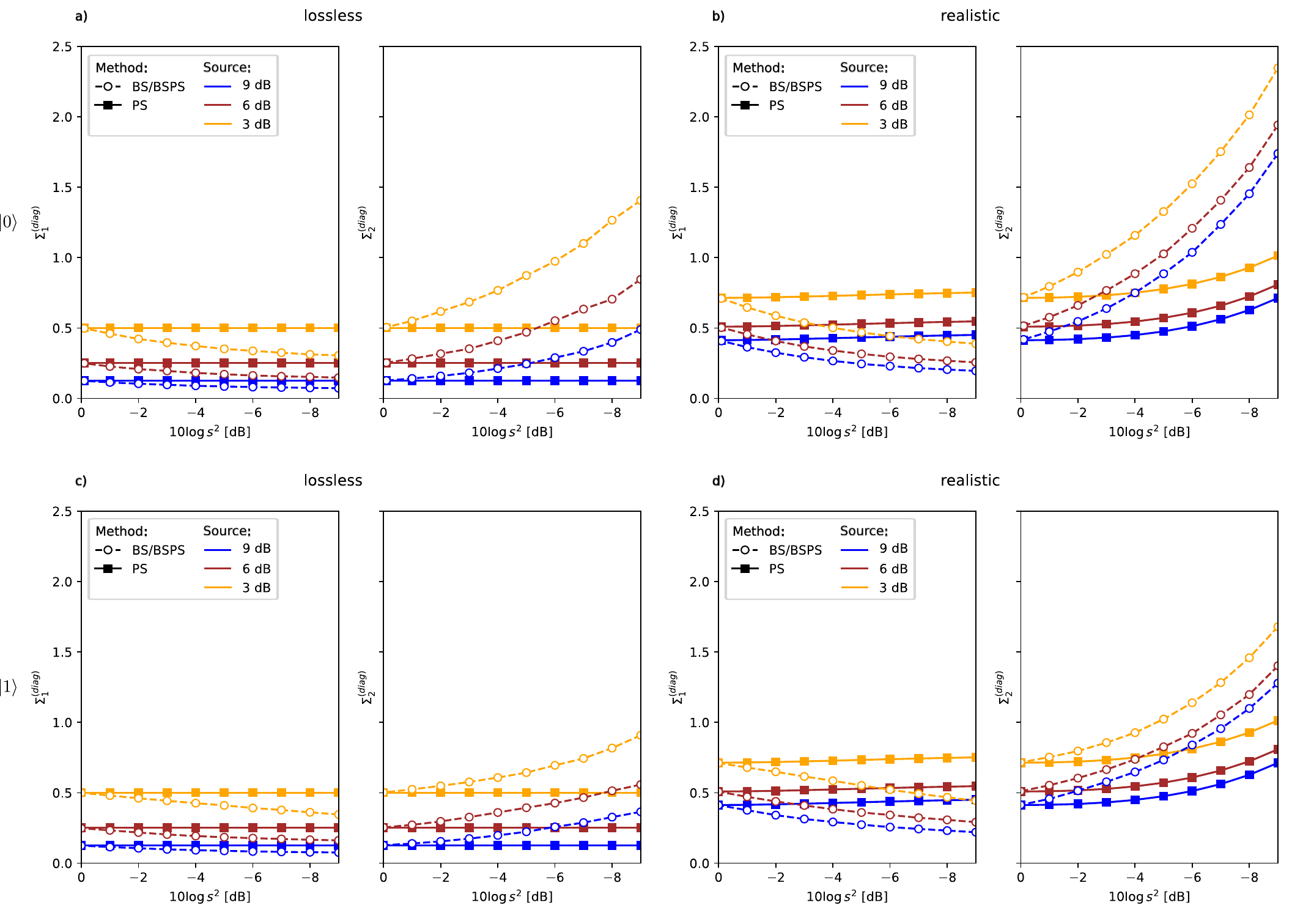}
\caption{\label{fig:var_supp} The variance of noise operators.  \textbf{a)}, and \textbf{b)} shows the variances for vacuum state. \textbf{c)}, and \textbf{d)} shows the variances for single photon state. Each figure is divided into the lossless case: $\eta_S =1$, $\eta_H = 1$, and the realistic case $\eta_S =0.8$, $\eta_H = 0.9$. The different lines distinguish different protocols, PS-sq (full lines with squares) and BS-sq (dashed lines with circles), and three different squeezing levels of the resource squeezed states, 9 dB (blue), 6 dB (brown), and 3 dB (yellow). The lines for BSPS-sq overlap with the lines for BS-sq.   } 
\end{figure*}
Note that the covariance of the teleported quadratures is not zero for $k \neq 0$ and $\eta_H < 1$ due to noise terms $\hat{N}_{x_{out}}$ and $\hat{N}_{p_{out}}$. This can be seen in Figure \ref{fig:cov}. The covariance for the input vacuum state is shown in the left figure. On the right is the covariance for the input single photon state. The covariance does not depend on the input state as expected. 
The noise matrix $\Sigma$ can be defined as in Eq. (13) in the main text. Due to generally non-zero off-diagonal elements, the matrix can be diagonalized: $\Sigma \rightarrow \Sigma^{\text{diag}}$. Its diagonal elements $\Sigma^{\text{diag}}_{1,2}$ can be plotted, see Fig \ref{fig:var_supp}. But the sum of these elements does not explain the behavior of fidelity - trace of the diagonalized matrix is not always lower for the BS-sq than for the PS-sq. Therefore, the re-parametrization given by Eq. (25) is needed.

\section{Numerical optimization of fidelity for BS-sq and BSPS-sq}\label{sec:appD}
\begin{figure*}[hpbt]
\includegraphics[width=0.9
\textwidth]{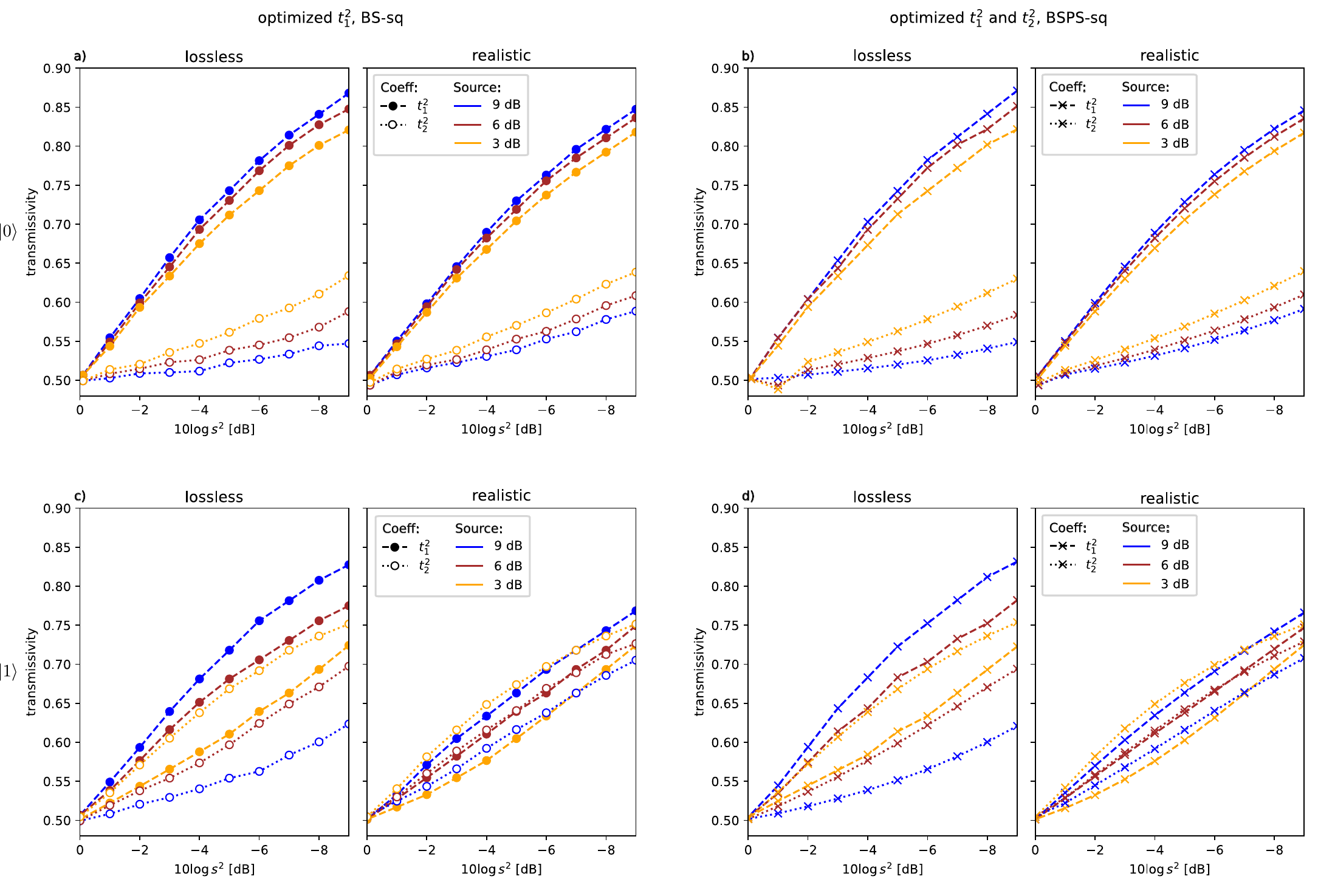}
\caption{\label{fig:t} The optimized $t_1^2$ and computed $t_2^2$ from Eq. \eqref{eq:g and k} for BS-sq in \textbf{a)} and \textbf{c)}. The optimized $t_1^2$ and $t_2^2$ for BSPS-sq in \textbf{b)} and \textbf{d)}. $\phi$ can be computed using Eq. \eqref{eq:g and k}.   Fig. \textbf{a)} and \textbf{b)} are for the vacuum state,  \textbf{c)} and \textbf{d)} are for the single photon state.   The left sub-figures always show the ideal case, $\eta_S =1$, $\eta_H = 1$. On the right, there are always realistic cases for them; it holds $\eta_S =0.8$, $\eta_H = 0.9$. The dashed lines marked with circles depict BS/BSPS-sq. The full line corresponds to PS-sq. The colors indicate the amount of source squeezing 9, 6, and 3 dB, in order blue, brown, and yellow.  } 
\end{figure*}

In this section, we show the parameters of variable beam splitters obtained from numerical optimization. To compare individual squeezers, we must fix the squeezing parameter $s$; therefore, we get the condition on optimization parameters. We optimize the parameters $t_1$ for the BS-sq. The parameter $t_2$ is given by the condition on $s$. We optimize $t_1$ together with $t_2$ for the BSPS-sq to achieve the highest possible fidelity. The parameter $\phi$ is given by the condition on $s$. The fidelity is defined in the main text. The BSPS-sq shows the same behavior as BS-sq because we get $\phi \approx 0$ in all cases. The squares of optimized parameters for given $s$ are shown in Fig. \ref{fig:t}. The left column of double-sub-figures shows optimized $t_1^2$ and given $t_2^2$ in the case of BS-sq. The right column shows optimized $t_1^2$ and $t_2^2$ for given $\phi$. The curves have quantitatively similar shapes. The minor inconsistencies are caused by different numerical approaches in optimizations: We compute the fidelity and choose the maximum value for every $t_1$ in the case of BS-sq. In the case of BSPS-sq, we used the differential evolution algorithm implemented in \textit{SciPy} \cite{scipy} to scan the parameter space of $t_1$ and $t_2$. The first row of double-sub-figures is for the vacuum input state, the second row is for the single photon input state. In every block, there are two sub-figures, each corresponding to lossless and realistic cases. 

\section{Numerical optimization of total noise for BS-sq}\label{sec:appE}
\begin{figure*}[hpbt]
\includegraphics[width=0.9
\textwidth]{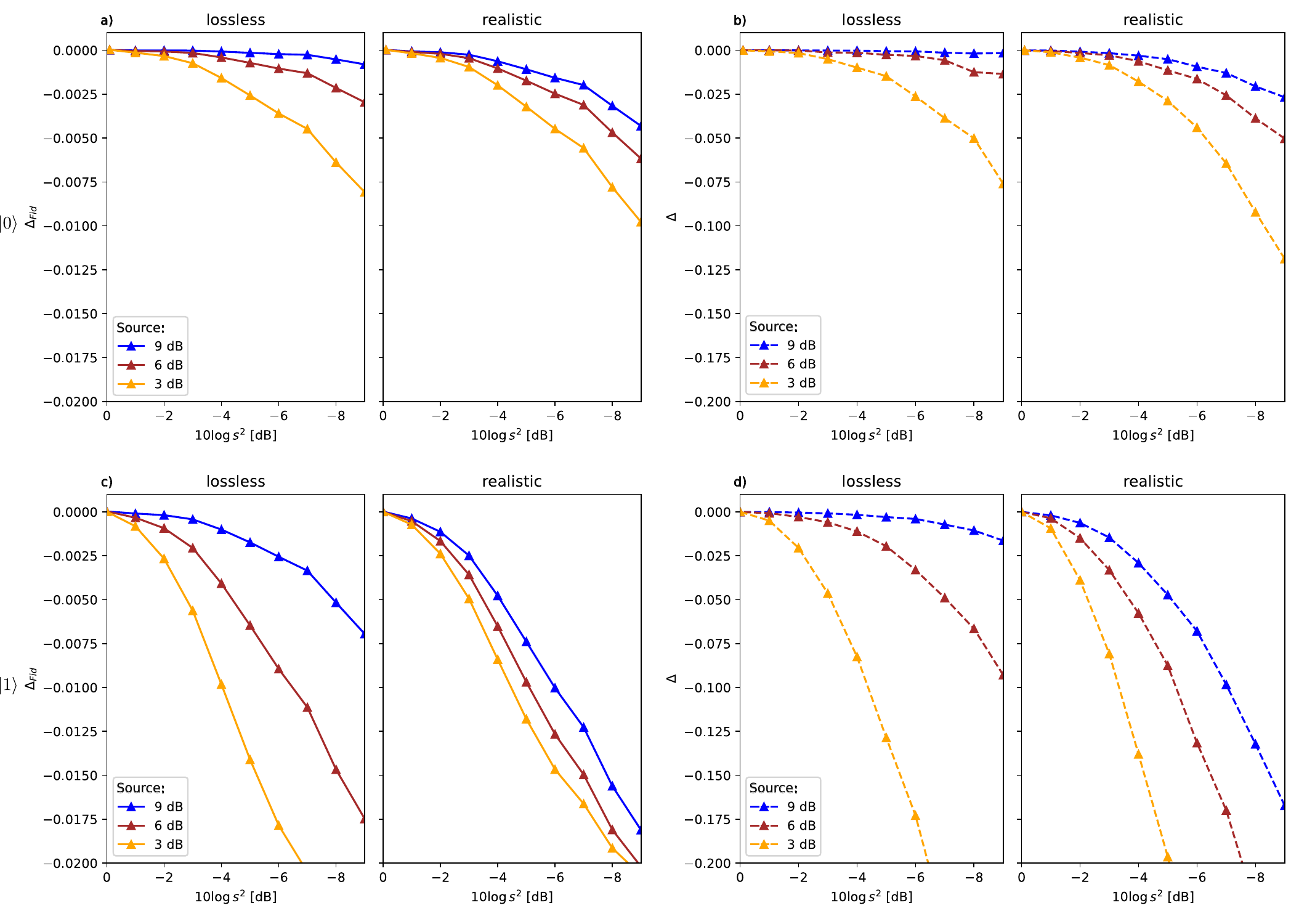}
\caption{\label{fig:Deltas} The differences of fidelities $\Delta_{Fid}$ and total noises $\Delta$ obtained from two different optimizations. Sub-figs. \textbf{a)} and \textbf{b)} represent differences for the input vacuum state, and see \textbf{c)} and \textbf{d)} are differences for the input single photon state. The colors indicate the amount of source squeezing 9, 6, and 3 dB, in order blue, brown, and yellow. } 
\end{figure*}

We also performed optimization of parameters $t_1$ and $t_2$ w.r.t. the minimum of total noise $N_T$. The total noise is defined in the main text.  We define
\begin{align}
    \Delta_{Fid} = F_{N_T} - F_{F}. 
\end{align}
Fidelity $F_{N_T}$ is the fidelity obtained from optimizing the parameters to a minimum of total noise, and fidelity $F_F$ is the fidelity obtained from optimizing parameters to maximal fidelity, as discussed in the main text.
We also define
\begin{align}
    \Delta = N_{T,N_T} - N_{T,F}. 
\end{align}
Total noise $N_{T,N_T}$ is the minimized total noise, and $N_{T,F}$ is the total noise obtained from the maximization of fidelity.
Fig. \ref{fig:Deltas} shows differences between fidelities $\Delta_{Fid}$ and between total noises $\Delta$.
Sub-figs. \textbf{a)} and  \textbf{b)} are the differences when vacuum state is injected. Sub-figs. \textbf{c)} and \textbf{d)} correspond to single photon input state. The negativity of differences implies that optimization w.r.t. total noise gives lower values of $N_{T,N_T}$ than $N_{T,F}$ and lower values of fidelity $F_{N_T}$ than $F_{F}$, which is the exact opposite of the case in Fig.~(2) and Fig.~(4) in the main text.  
Therefore, the behavior of total noise does not immediately explain the behavior of the fidelity in the sense that the less total noise, the higher the fidelity, mainly in non-Gaussian states.  Due to a small absolute value of difference in fidelities $\Delta_{Fid}$, the fidelities $F_{N_T}$ and $F_F$ have similar behavior. Therefore, the optimization to minimize total added noise performs similarly to the maximization of fidelity.

\section{Details on BAS-squeezer}\label{sec:appF}
\begin{figure*}[hpbt]
\includegraphics[width=0.9
\textwidth]{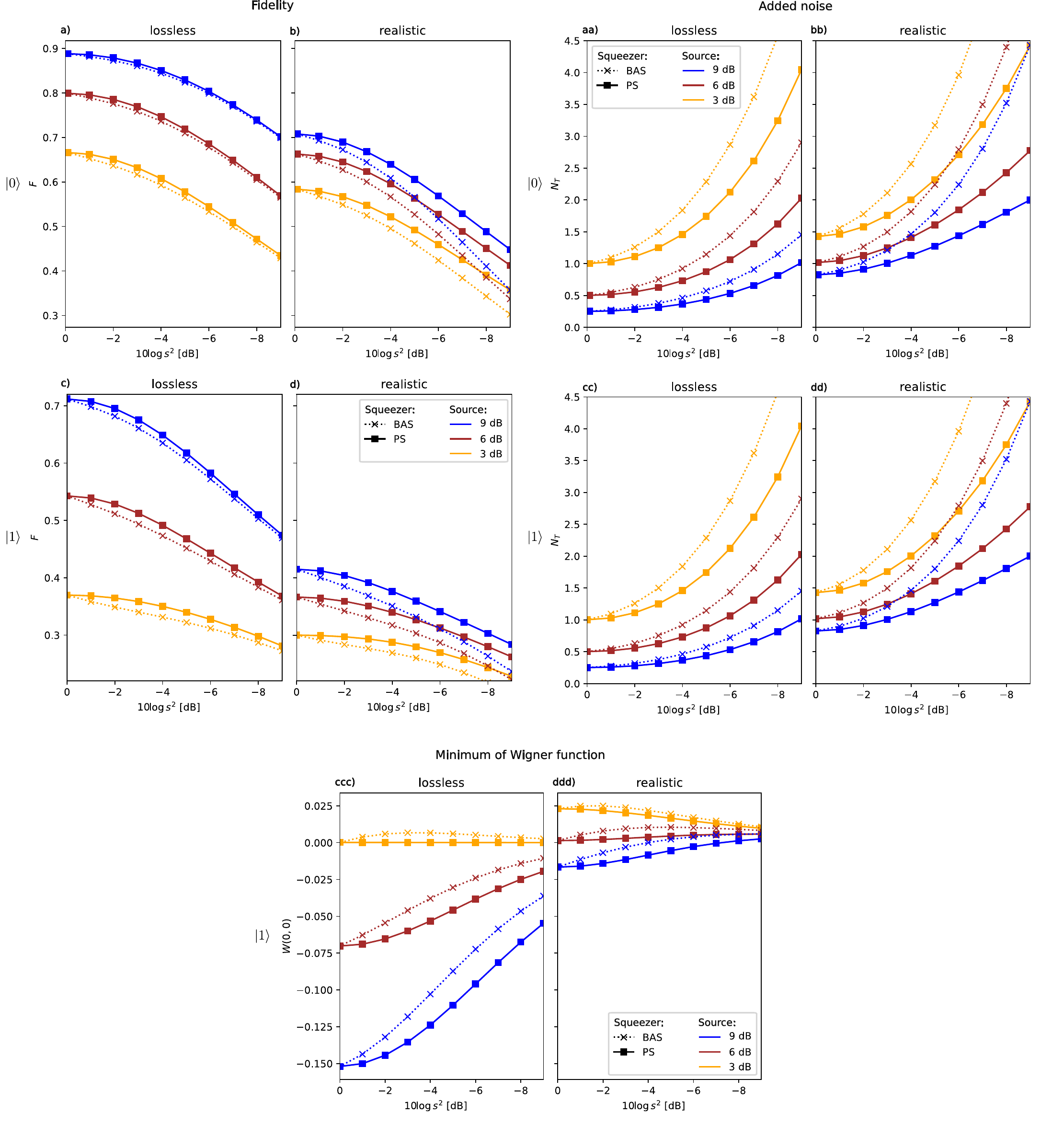}
\caption{\label{fig:bas_sq} The comparison of fidelities, added noise, and minimum of the Wigner function for BAS-sq. \textbf{a)} and \text{aa)} Vacuum states in the lossless scenario; \textbf{b)} and \textbf{bb)} vacuum states in the realistic scenario with $\eta_S = 0.8$ and $\eta_H = 0.9$; \textbf{c)}, \textbf{cc)} and \textbf{ccc)} Single photon states in the lossless scenario; \textbf{d)}, \textbf{dd)} and \textbf{ddd)} Single photon states in the realistic scenario with $\eta_S = 0.8$ and $\eta_H = 0.9$. The different lines distinguish different protocols,  PS-sq (full lines with squares) and  BAS-sq (dotted lines with crosses), and three different squeezing levels of the resource squeezed states, 9 dB (blue), 6 dB (brown), and 3 dB (yellow).  } 
\end{figure*}

The BS-squeezer has only one free parameter $t_0$, which comes from the unbalanced beam splitter in measurement-induced squeezing \cite{filip_marek}. The output state is squeezed input with additional noise $ \hat{N}_{x_{bas}}$ and $\hat{N}_{p_{bas}}$. This noise is added to the noise from the teleportation gate itself. Naturally, this leads to even worse performance in fidelity and in the minimum of the Wigner function, see Fig~\ref{fig:bas_sq}. The poor performance of BAS-sq is caused by the additional resource of squeezing and an additional homodyne detector, both of which contribute to additional noise in the output state. We compare BAS-sq with PS-sq in fidelity, at the minimum of the Wigner function, and at the total added noise.

\section{Details on Genuine non-Gaussianity}\label{sec:appG}
\begin{figure*}[hpbt]
\includegraphics[width=0.9
\textwidth]{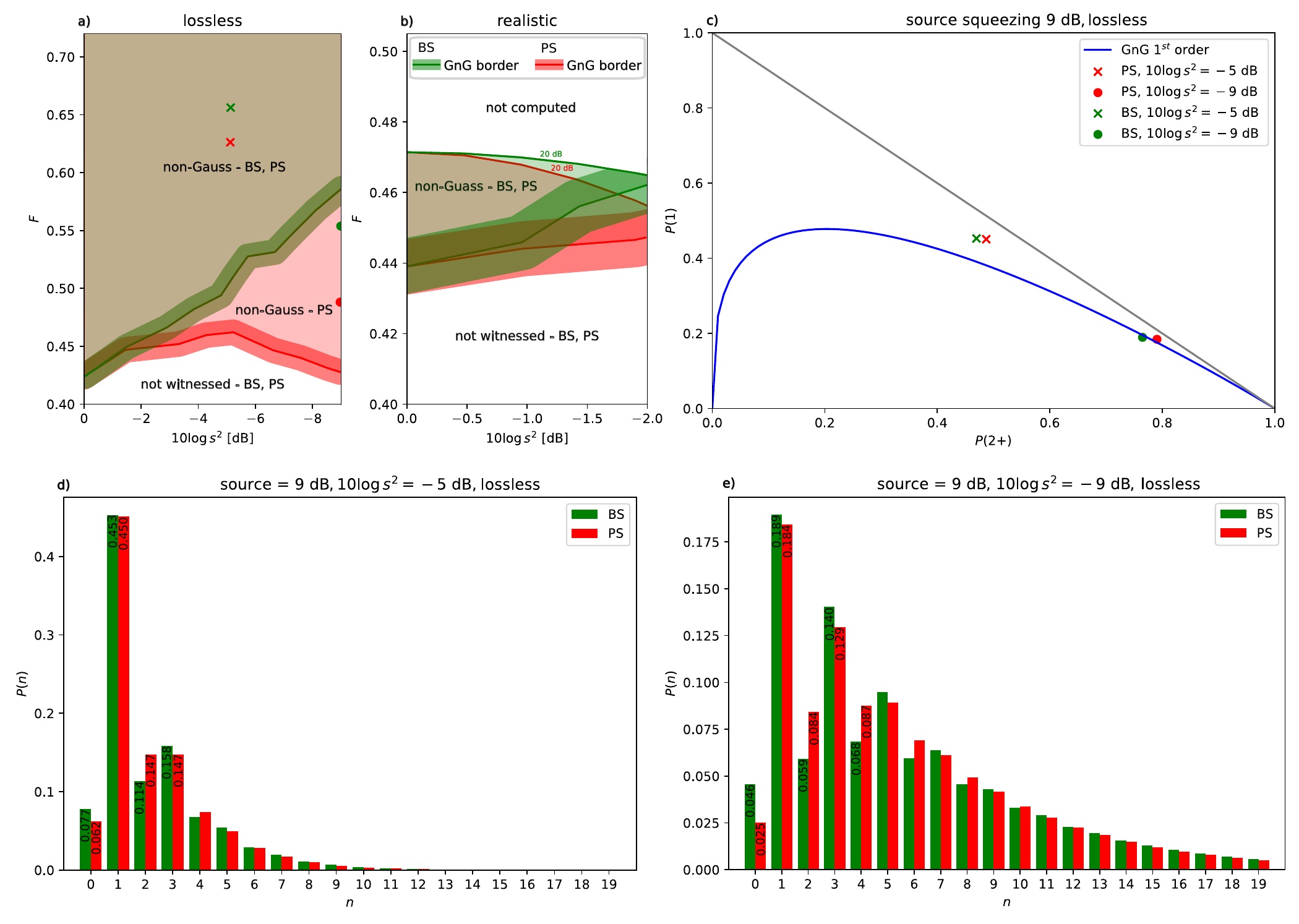} 
\caption{\label{fig:gng_supp}
 The comparison of GnG regimes is shown for fidelity. \textbf{a)} Lossless scenario; \textbf{b)} realistic scenario with $\eta_S = 0.8$ and $\eta_H = 0.9$. The different areas distinguish GnG regimes, PS-sq and non-Gauss regime (light red), PS-sq and not witnessed regime (white), BS-sq and non-Gauss regime (light green), BS-sq and not witnessed regime (white). The thick borderline indicates the iterated GnG boundary for every given resource squeezing. The thickness denotes numerical precision. In the realistic case, there is a third type of region - not computed. This area requires non-trivial computational power.  The colored crosses and circles denote a specific states of 9 dB of source squeezing: crosses - $10\log s^2 = -5$ dB, circles - $10\log s^2 = -9$ dB. The color indicates affiliation to the squeezer type: BS-sq - green, PS-sq - red.  Their photo-statistics are shown in \textbf{d)} and \textbf{e)}. Their certification of GnG witness is shown on \textbf{c)}. The blue curve is the 1st order of Genuine non-Gaussianity, and it was first computed in \cite{lachman}. } 
\end{figure*}

Fig. \ref{fig:gng_supp} and sub-figures \textbf{a)} and \textbf{b)} show the fidelities of single photon states squeezed by the BS and PS squeezers with continuous levels of resource squeezing. The fidelities are divided by the GnG witness - the GnG border, and two regions are created: non-Gaussian and not witnessed. The thickness of the border denotes numerical precision.  When the specific state with the specific fidelity is witnessed by the first order of GnG \cite{lachman}, it is in the non-Gaussian region. The example can be seen in \textbf{c)}. We take four states with a common source squeezing of 9 dB. The differences between them are given by the squeezing parameter $s$: $10 \log s^2 = $ -5 dB, -9 dB, and by the squeezing method (squeezer). These states have corresponding fidelity on the right side of sub-figure \textbf{a)}. The red cross and red circle correspond to the PS-squeezer. The green cross and circle represent BS-squeezer.  While BS-sq offers higher fidelities for given $s$, it also leads to quicker loss of non-Gaussianity certified by GnG. This implies that the squeezed states produced by BS-sq contain higher contributions of the vacuum state in their photo-statistics, as can be seen in sub-figures \textbf{d)} and \textbf{e)}. Two photo-statistics are plotted. Sub-fig. \textbf{d)} shows the photo-statistics of the states marked with crosses (higher values of the squeezing parameter $s$).  Sub-fig. \textbf{e)} shows the photo-statistics of the states marked with circles (lower values of the squeezing parameter). However, from the analysis of the minimum of the Wigner function, we know that BS-sq generally outperforms PS-sq,  but PS-sq outperforms BS-sq in non-Gaussianity certified by GnG. This indicates that the GnG could be improved when the distribution of noise is taken into account.  In the realistic case of sub-fig \textbf{b)}, there is a third area that is out of computational power - not computed. This area is limited by the fidelity curve computed for 20 dB of resource squeezing. In fact, the same value of source squeezing was used in the lossless case, but in sub-Fig. \textbf{a)} there is a cut at $F = 0.72$. We conclude: If one wants to use the GnG metric to certify the non-Gaussianity of the output squeezed single photon state, the BS-squeezer has no advantage. In the following research, the larger subset of non-Gaussian states should be considered before any general conclusion regarding the GnG metric in the case of teleportation-based squeezers.

\end{document}